\documentstyle[12pt,aaspp4]{article}
\begin{document}
\title{POWER SPECTRUM OF VELOCITY FLUCTUATIONS
IN THE UNIVERSE}
\author {MIRT GRAMANN}
\affil{Tartu Astrophysical Observatory, T\~oravere EE-2444, 
Estonia}

\begin{abstract}
We investigate the power spectrum of velocity fluctuations 
in the universe starting from four different measures of 
velocity: the power spectrum of velocity fluctuations from peculiar 
velocities of galaxies; the rms peculiar velocity of galaxy 
clusters; the power spectrum of velocity fluctuations from 
the power spectrum of density fluctuations in the galaxy distribution; and 
the bulk velocity from peculiar velocities of galaxies.
There are various way of interpreting the observational data:

(1) The power spectrum of velocity fluctuations 
follows a power law, $V^2(k) \sim k^2$, on large scales, achieves a 
maximum $V(k) \sim 500$ km s$^{-1}$ at a wavelength $\lambda \sim
120h^{-1}$ Mpc, and declines as $V^2(k) \propto k^{-0.8}$
on small scales. This type of power spectrum is predicted by a mixed 
dark matter model with density parameter $\Omega_0=1$. This model is 
consistent with all data observed, except the rms peculiar 
velocity of galaxy clusters. 

(2) The shape of the power spectrum of velocity fluctuations is similar
to that in model (1), but the amplitude is lower ($\sim 300$ km s$^{-1}$
at $\lambda \sim 120h^{-1}$ Mpc). This power spectrum is predicted 
by a low-density cold dark matter model with density parameter 
$\Omega_0 \simeq 0.3$. 

(3) There is a peak in the power spectrum of
velocity fluctuations at a wavelength $\lambda \simeq 120h^{-1}$ Mpc and 
on larger scales the power spectrum decreases with an index 
$n\simeq 1.0$. The maximum value of the function $V(k)$ is 
$\sim 420$ km s$^{-1}$. This power spectrum is consistent with the power 
spectrum of the galaxy distribution in the Stromlo-APM redshift survey 
provided the parameter $\beta$ is in the range $0.5-0.6$. 

(4) There is a peak in the power spectrum as in model (3), but on 
larger scales the amplitude of fluctuations is higher than that estimated 
starting from the observed power spectrum of galaxies. For the parameter 
$\beta$ in the range $0.4-0.5$, the observed rms cluster 
peculiar velocity is consistent with the rms amplitude of the 
bulk flow $\sim 340$ km s$^{-1}$ at the scale $60h^{-1}$ Mpc. 
In this case the value of the function $V(k)$ at wavelength 
$\lambda=120h^{-1}$ Mpc is $\sim 350$ km s$^{-1}$.

In the future, larger redshift surveys and more 
accurate observations of peculiar velocities of galaxies and clusters 
will help to constrain the power spectrum of velocity fluctuations 
in the universe.
\end{abstract}


\section{INTRODUCTION}

The velocity of matter in the universe, ${\bf u} ({\bf r})$,
can be expressed as a sum of the mean Hubble expansion velocity 
${\bf v}_{\rm H} = H_0{\bf r}$ and a field of velocity fluctuations
$$
{\bf v} \, ({\bf r}) \equiv \ {\bf u} ({\bf r}) - H_0 {\bf r},
\eqno(1)
$$
where $H_0$ is the Hubble constant.  
The peculiar velocity field ${\bf v}({\bf r)}$ in the volume $V_u$ 
can be expressed in terms of its Fourier components 
$$
{\bf v} \, ({\bf r}) = {V_u^{ 1/2} \over (2 \pi)^{3/2} }
\int {\bf v}_{\bf k} \,\, \exp(i {\bf k} {\bf r}) \,\, d^3 k,
\eqno(2)
$$
and quantified in terms of the power spectrum 
$P_{\rm v} (k) \equiv < \! \vert v_{{\bf k}{\rm x}}\vert^2 
+\vert v_{{\bf k}{\rm y}}\vert^2 + \vert v_{{\bf k}{\rm z}}\vert^2 \! >$. 
If the field 
${\bf v} \, ({\bf r})$ is an isotropic Gaussian field, 
then the different Fourier components are uncorrelated, 
and the power spectrum provides a complete statistical 
description of the field.

Previous studies in the literature have investigated the density 
fluctuations and the field of velocity fluctuations in real space
(see Dekel 1994, Strauss \& Willick 1995 for a review). 
This paper, however, concentrates on the power spectrum of velocity 
fluctuations. We will describe the velocity spectrum by
$$
V^2 (k) \equiv {1 \over 2 \pi^2} \, k^3 P_{\rm v}(k) . 
\eqno(3)
$$
The function $V^2 (k)$ gives the contribution to the velocity 
dispersion per unit interval in $\ln k$,
$$
< \! v^2 \! > \equiv {1 \over V_u} \int v^2({\bf r}) \,\, d^3 r 
={1 \over 2 \pi^2} \int P_{\rm v}(k) k^2 dk 
= \int V^2 (k) \,\, {dk \over k} \,.
\eqno (4)
$$
The rms velocity fluctuation on a given scale $r$ can be expressed as
$$
< \! v^2 \! \, (r) > = \int V^2 (k) \, W^2(kr) \,\, {dk \over k} \, ,
\eqno (5)
$$
where $W(kr)$ is the Fourier transform of the window function applied
to determine the peculiar velocity field. For a Gaussian window 
function, the rms velocity of matter is given by
$$
v_{\rm rms}^2 (r) = \int V^2 (k) \exp(-r^2 \,k^2) \,\, {dk \over k} \,.
\eqno (6)
$$

We can study the rms velocity of matter in the universe using 
clusters of galaxies as tracers. Bahcall, Gramann, \& Cen (1994) and 
Gramann et al. (1995) compared the motions of clusters of galaxies 
with the motion of the underlying matter distribution in different 
cosmological models. The rms cluster peculiar velocity is similar 
to the rms peculiar velocity of matter smoothed with a 
Gaussian window of radius $r \simeq 3h^{-1}$ Mpc. The observed 
peculiar velocity function of galaxy clusters was investigated 
by Bahcall \& Oh (1996). They found an rms one-dimensional cluster 
peculiar velocity $< \! v^2_{\rm 1D} \! >^{1/2} =293 \pm 28$ km s$^{-1}$. 
This corresponds to a three-dimensional rms velocity 
$< \! v^2 \! >^{1/2} =507 \pm 48$ km s$^{-1}$.

What is the origin of the velocity dispersion of galaxies and 
clusters of galaxies? Does the velocity dispersion of galaxy systems 
originate mostly from the small-scale velocity fluctuations of matter
with wavelengths $\lambda <100h^{-1}$ Mpc, or from the large-scale
velocity fluctuations with wavelengths $\lambda >100h^{-1}$ Mpc? Or is 
there a peak in the function $V^2(k)$ at $\lambda \sim 100h^{-1}$ Mpc that 
contributes most to the velocity dispersion?

We will examine the power spectrum of velocity fluctuations
and rms velocity of matter starting from the power spectrum of 
density fluctuations derived from large galaxy surveys. 
In the linear approximation, the continuity equation yields a
relation between the density contrast $\delta$ and the 
peculiar velocity, 
$$
\nabla {\bf\cdot} {\bf v} = - f(\Omega_0) \, H_0 \, \delta \,,
\eqno(7)
$$
where the function $f(\Omega_0)$ is the linear velocity growth factor 
and $\Omega_0$ is the cosmological density parameter at the present 
moment. The function $f(\Omega_0) \approx \Omega_0^{\,0.6}$ 
(Peebles 1980). In Fourier space equation (7) takes the form 
$$
{\bf v}_{\bf k} {\bf\cdot}  i {\bf k} = - f(\Omega_0) \, H_0 \, 
\delta_{\bf k},
\eqno (8)
$$
where $\delta_{\bf k}$ is the Fourier transform of the 
density field. The linear growing mode is irrotational. If the 
velocity field has no vorticity, the function $V^2(k)$ can be 
determined as 
$$
V^2 (k) = {1 \over 2 \pi^2} \,\, f^2(\Omega_0) H_0^2 \,\,  k P(k) ,
\eqno (9)
$$
where $P(k) \equiv < \! \vert\delta_{\bf k}\vert^2 \!>$ is the 
power spectrum of density fluctuations. 

The power spectrum of density fluctuations in the mass distribution
has been estimated by Kolatt \& Dekel (1997), via the use of galaxy 
peculiar velocities. In this paper we will derive the power spectrum 
of velocity fluctuations on the basis 
of their results. We find that the power spectrum estimated by 
Kolatt and Dekel (1997) corresponds to an rms velocity 
$< \! v^2 \! >^{1/2} \simeq 700$ km s$^{-1}$ for the matter 
distribution smoothed on scales $\sim 3h^{-1}$ Mpc. This value is larger 
than that observed by Bahcall \& Oh (1996) for the rms cluster peculiar 
velocity. Therefore, either the power spectrum of density fluctuations 
estimated by Kolatt \& Dekel (1997) is overestimated or the 
rms cluster peculiar velocity determined by Bahcall and Oh (1996) 
is underestimated. Available data are insufficient 
to distinguish between these scenarios and so we must consider 
both possibilities.

We will examine the power spectrum of velocity fluctuations
starting from the power spectrum of density fluctuations derived 
from large redshift surveys of galaxies. The power spectrum of the 
galaxy distribution has been measured from a number of large galaxy 
surveys. In this paper we will 
investigate the peculiar velocity field in the Stromlo-APM and 
Las Campanas redshift surveys (Tadros \& Efstathiou 1996; 
Lin et al. 1996). The amplitude of the velocity fluctuations 
derived from the galaxy distribution depends on the parameter 
$\beta=f(\Omega_0)/b$, where $b$ is the bias factor for the galaxies. 
We will estimate the parameter $\beta$ on the basis of the 
observed rms cluster peculiar velocity. 

The power spectrum of the galaxy distribution in the Stromlo-APM 
redshift survey peaks at a wavenumber $k=0.052h$ Mpc$^{-1}$ 
(or at a wavelength $\lambda = 120h^{-1}$ Mpc). Available data,
however, are insufficient to say whether the peak in the Stromlo-APM
survey reflects a real feature in the galaxy distribution. It is
likely that the decline in the power spectrum at wavenumbers $k \le
0.052h$ Mpc$^{-1}$ is partly due to the effects
of the uncertainty in the mean number density of optical galaxies (see
Tadros \& Efstathiou 1996 for discussion of this effect).
Einasto et al. (1997) found a well-defined peak at the same wavelength, 
$\lambda = 120h^{-1}$ Mpc, in the power spectrum of galaxy clusters.  
On the other hand, there is no well-defined peak in the three-dimensional 
power spectrum of the galaxy distribution in the
Las Campanas redshift survey derived by Lin et al. (1996). 
There is a striking peak at $\lambda \approx 100h^{-1}$ Mpc in the
two-dimensional power spectrum of the Las Campanas redshift survey 
(Landay et al. 1996). A similar peak at $128h^{-1}$ Mpc in the
one-dimensional power spectrum of a deep pencil beam survey was
detected by Broadhurst et al. (1990). If there is an excess of power
in the universe around a scale of $120h^{-1}$ Mpc, then this scale contributes 
most to the velocity dispersion of galaxy systems. 
We will investigate the relation between the velocity dispersion
on scales $\sim 3h^{-1}$ Mpc and maximum value for the power spectrum
of velocity fluctuations on wavelengths $\sim 120h^{-1}$ Mpc.

To characterize the large-scale peculiar velocity field we can 
use the bulk velocity of galaxies. The bulk velocity averaged over 
spheres of radius $r$ is determined as
$$
v_b^2(r)=9 \, \int V^2 (k) \,
{(\sin kr - kr \cos kr)^2 \over (kr)^6}  \,\, {dk \over k} \,.
\eqno(10)
$$
The bulk velocity of galaxies on $\sim 60h^{-1}$ Mpc scales is determined 
by the amplitude of the density and velocity fluctuations in the universe 
on scales with wavenumber $k\le 0.05h$ Mpc$^{-1}$ 
($\lambda \ge 120h^{-1}$ Mpc). 
We will estimate the bulk velocity starting from the power 
spectrum of the galaxy distribution.

In the linear approximation the power spectrum of velocity fluctuations 
is directly related to the power spectrum of density fluctuations.
Formally these power spectra are identical in their information 
content. Consequently one may ask why it is necessary to investigate
the velocity power spectrum at all.
The properties of the peculiar velocity field, however,  
are best visualized and understood in terms of the velocity spectrum, 
just as the properties of the density field are best expressed in terms of 
the density spectrum. For instance, the quantities discussed in this paper 
(velocity dispersion and bulk velocity) are easily derived from the 
velocity power spectrum. It is therefore advantageous to combine both 
approaches to get a better understanding of the large-scale matter 
distribution in the universe.

The paper is organized as follows. In \S 2 we estimate the
power spectrum of velocity fluctuations from peculiar velocities
of galaxies and analyze the rms velocity of matter in the universe 
in more detail. In \S 3 we analyze the power spectrum of the
galaxy distribution measured from various redshift surveys and in \S 4 present
a method for estimating the power spectrum of velocity fluctuations 
and rms velocity of matter starting from the power spectrum of 
the distribution of galaxies. In \S 5 we examine the power 
spectrum of velocity fluctuations in the Las Campanas redshift 
survey and in \S 6 we investigate the peculiar velocity 
fluctuations in the Stromlo-APM  redshift survey. The discussion 
and summary are presented in \S 7.

A Hubble constant of $H_0=100h$ km s$^{-1}$Mpc$^{-1}$ is used 
throughout this paper. 

\section{PECULIAR VELOCITIES OF GALAXIES AND CLUSTERS OF GALAXIES}

Kolatt \& Dekel (1997, hereafter KD) derived the power spectrum of 
density fluctuations from the Mark III catalog of peculiar velocities 
(Willick et al. 1997). This catalog consists of more than 3000 galaxies 
from several different data sets of spiral and elliptical/SO galaxies
with distances estimated by the Tully-Fisher and D$_n$ - $\sigma$ methods.
The fractional error in the distance to each galaxy is of the order 17
-- 21\%.

KD used the POTENT method to recover the smoothed three-dimensional 
velocity field from the observed radial velocities (Bertschinger 
et al. 1990). The method assumes that the velocity field is potential. 
The velocity field was smoothed with a Gaussian of radius $12h^{-1}$ Mpc, 
and then the density field was computed using a quasi-linear 
solution for the continuity equation. This approximation 
reduces to relation (7) in the linear regime. KD applied an empirical
correction procedure to recover the true power spectrum from the 
observed power spectrum of density fluctuations. This correction
procedure was based
on mock catalogs designed to mimic the observational data.

KD derived values for the function $f^2(\Omega_0)P(k)$ with 
$1\sigma$ errors. Figure~1a shows the rms amplitude of velocity 
fluctuations, $V(k)$, computed on the basis of their results, using 
equation (9). The function $V(k)$ has been calculated for the
wavenumber range 
$0.061 < k < 0.172h$ Mpc$^{-1}$. For the wavenumbers 
$k=0.172h$ Mpc$^{-1}$, $k=0.102h$ Mpc$^{-1}$ and $k=0.061h$ Mpc$^{-1}$, 
the rms amplitude of velocity fluctuations $V(k) = 414 \pm 52$ km/s, 
$V(k)=489 \pm 66$ km/s and $V (k) = 502 \pm 96$ km/s, respectively.
To describe the power spectrum of velocity fluctuations for the 
peculiar velocity data we can use the fitting function
$$
V^2(k)=2V^2 (k_0) \, (k/k_0)^{n+1} \,\, [1+ (k/k_0)^{n+m}]^{-1} \,
\eqno(11)
$$
where $k_0=0.052h$ Mpc$^{-1}$, $V(k_0)=496$ km/s, $n=1$ and $m=1.85$.
This function is consistent with the data at a confidence level of
$>99$\% (based on a $\chi^2$ test).

We have estimated the rms peculiar velocity of matter corresponding 
to the power spectrum estimated by KD. The rms peculiar
velocity was computed using equation (6).
Figure~1b shows the rms peculiar velocity for the matter distribution
at radii $r=1h^{-1}$ Mpc to $r=5h^{-1}$ Mpc for the
velocity spectrum (11). At a smoothing radius $r=3h^{-1}$ Mpc the rms 
peculiar velocity $v_{\rm rms}=709$ km s$^{-1}$.

The rms peculiar velocity calculated using approximation (11)
can be considered as a lower limit for the power spectrum derived 
by KD. We have assumed that on scales with wavenumber $k<0.06h$ Mpc$^{-1}$ 
($\lambda >100h^{-1}$ Mpc) the velocity spectrum decreases 
monotonically. If there is a peak in
the velocity spectrum on scales with wavenumber $k \sim 0.05h$ Mpc$^{-1}$ or 
if the turnover in the spectrum occurs at larger scales, then
the rms peculiar velocity would be higher than the value computed
using approximation (11). Therefore, the power spectrum 
estimated by KD corresponds to an rms peculiar velocity 
which is larger than $700$ km s$^{-1}$ for the matter 
distribution on scales $\sim 3h^{-1}$ Mpc.

For comparison we show in Figure~1b the rms cluster peculiar velocity
found by Bahcall \& Oh (1996). They determined the 
peculiar velocity function of galaxy clusters using an accurate 
sample of peculiar velocities of clusters obtained by 
Giovanelli et al. (1996). This sample consisted of 22 clusters and 
groups of galaxies with peculiar velocities based on Tully-Fisher 
distances to Sc galaxies. Bahcall \& Oh (1996) found an 
rms one-dimensional cluster peculiar velocity of 
$293 \pm 28$ km s$^{-1}$, which corresponds to a 
three-dimensional rms velocity of $507 \pm 48$ km s$^{-1}$. 

Numerical simulations show that the velocity distribution of clusters
is similar to that of the matter when the matter distribution is smoothed
with a Gaussian of radius $3h^{-1}$ Mpc (Bahcall, Gramann, \& Cen 1994,
Gramann et al. 1995). The velocity distribution of the unsmoothed matter
exhibits higher velocities than the clusters, especially for 
$\Omega=1$ models, due to the high velocity dispersion
of matter within clusters. The $3h^{-1}$ Mpc smoothing of the matter
distribution eliminates this nonlinear effect. The rms cluster
peculiar velocity is similar to, or somewhat higher than, the rms peculiar
velocity of the smoothed matter. Therefore, the rms cluster 
velocity determines an upper limit for the rms velocity 
of matter on scales $\sim 3h^{-1}$ Mpc. 

The power spectrum of velocity fluctuations estimated from the mass
power spectrum of Kolatt \& Dekel (1997) corresponds to an rms peculiar velocity 
which is larger than expected on the basis of the observed rms 
cluster peculiar velocity determined by Bahcall \& Oh (1996). 
Given the large errors associated with
peculiar velocity measurements of galaxies, this discrepancy 
is not very large. The sample used by Bahcall \& Oh (1996) 
consisted of only 22 clusters; a small sample can introduce significant 
statistical uncertainties. On the other hand, the peculiar 
velocities of galaxies used to estimate the power spectrum are 
contaminated by distance errors, as well as being sparsely and
non-uniformly sampled. The systematic errors may be more 
complicated than envisaged by Kolatt \& Dekel (1997). 

The three-dimensional rms cluster peculiar velocity 
$\sim 500$ km s$^{-1}$ is in reasonable agreement with the 
results of Marzke et al. (1995), who studied the rms relative 
peculiar velocity between galaxy pairs 
separated by $\sim 1 h^{-1}$ Mpc. They found an rms one-dimensional 
velocity $\sigma_{12}=540 \pm 180$ km s$^{-1}$ from an analysis of 
the CfA-2 and Southern Sky redshift surveys. Assuming the velocities 
of the galaxies are isotropic and independent, a
three-dimensional rms velocity $v_{rms}=500$ km s$^{-1}$ corresponds to a 
pairwise rms velocity $\sigma_{12} = (2/3)^{1/2} \, v_{\rm rms} = 408$ km s$^{-1}$. However, the rms velocity of galaxies is probably higher than the rms 
velocity of clusters because of the high velocity dispersion within
the clusters. Also, the velocities of galaxy pairs with separation
$\sim 1h^{-1}$ Mpc are correlated. 
Together, these effects can easily explain the difference between the value $408$ km s$^{-1}$, predicted on the basis of the cluster rms velocity, and the measured value of $540$ km s$^{-1}$.

\section{POWER SPECTRUM OF THE GALAXY DISTRIBUTION}

Let us now consider the power spectrum of the galaxy distribution 
determined from different redshift surveys. The power spectrum of 
the galaxy distribution in the Stromlo-APM 
redshift survey has been computed by Tadros \& Efstathiou (1996, 
hereafter TE). The Stromlo-APM redshift survey is a 
1 in 20 sparsely sampled subset of 1787 galaxies selected 
from the APM Galaxy survey (Maddox et al. 1990). The survey is 
described in detail by Loveday et al. (1992). The 
median redshift of the Stromlo-APM survey is $z=0.05$.
TE estimated the power spectrum of density fluctuations
using different volume limited and flux limited samples. They 
found that galaxy density power spectra are insensitive to 
the volume limit as well as to the weights applied in the 
analysis of flux limited samples. They also tested the power spectrum
estimator against simulated galaxy catalogues 
constructed from N-body simulations and showed that the methods 
applied provide nearly unbiased estimates of the power spectrum at 
wavenumbers $k > 0.04h$ Mpc$^{-1}$. At smaller wavenumbers the 
power spectrum may be underestimated. 

Figure 2 shows the power spectrum of galaxy clustering, 
$P^s_{gal}(k)$, in the Stromlo-APM redshift survey with $1\sigma$ 
errors derived by TE. We present estimates for the flux-limited 
sample with $P(k)=8000h^{-3}$ Mpc$^3$ in the weighting function
(see TE for details). The power spectrum of the galaxy distribution
peaks at the wavenumber $k_0=0.052h$ Mpc$^{-1}$ 
($\lambda = 120h^{-1}$ Mpc). To describe the power spectrum in 
the Stromlo-APM survey we can use the fitting function
$$
P(k)=\cases {P(k_0)(k/k_0)^n, &if $k<k_0$;\cr   
P(k_0) (k/k_0)^m, &if $k > k_0$ , \cr}
\eqno(12)
$$
where $k_0=0.052h$ Mpc$^{-1}$, $P(k_0)=3.16 \cdot 10^4 h^{-3}$ 
Mpc$^{3}$, $n=0.5$ and $m=-2$. This function is consistent with the
power spectrum in the Stromlo-APM survey at a confidence level 
of $\sim 70$\%. 

The Las Campanas redshift survey contains 23,697 galaxies, with an 
average redshift z=0.1, distributed over six slices in the north and 
south Galactic caps (Shectman et al. 1996). Figure~2 shows the 
three-dimensional power spectrum of galaxy clustering computed by 
Lin et al. (1996). The observed power spectrum of galaxy clustering 
in the Las Campanas survey can be fit by
$$
P(k)=2 P(k_0)\, (k/k_0)^n \,\, [1+(k/k_0)^{m+n}]^{-1} ,
\eqno(13)
$$
where $k_0=0.06h^{-1}$ Mpc, $P(k_0)=1.28 \cdot 10^4h^{-3}$ Mpc$^{3}$,
$n=1.2$ and $m=1.8$. The function (13) is consistent with the 
power spectrum in the Las Campanas survey at a confidence level of 
$>99$\%. Figure~2 also shows the power spectrum of 
the galaxy distribution in the SSRS2+CfA2 redshift survey 
determined by da Costa et al. (1994). The power spectrum 
is presented for a volume limited sample with a distance limit 
$r=101h^{-1}$ Mpc.

At wavenumbers $k \geq 0.06h$ Mpc$^{-1}$ ($\lambda <100h^{-1}$ Mpc) 
the power spectra from the different redshift surveys are consistent. 
On larger scales the power spectrum of the 
galaxy distribution is relatively poorly constrained by observations.
At wavenumbers $k \simeq 0.04 - 0.06h$ Mpc$^{-1}$ the power 
spectrum of the galaxy distribution in the Stromlo-APM survey is  a
factor of two higher than the power spectrum derived from the 
Las Campanas survey. There is no well defined peak in the 
three-dimensional power spectrum derived by Lin et al. (1996). 
Landay et al. (1996) measured the two-dimensional power spectrum 
of the Las Campanas survey and found a strong peak in the 
power spectrum at $\sim 100h^{-1}$ Mpc. The signal was detected 
in two independent directions on the sky and identified with 
numerous structures visible in the survey which appear as walls 
and voids. Given the geometry of the Las Campanas survey the 
three-dimensional analysis is not as sensitive as the two-dimensional
analysis to structures on scales $>50h^{-1}$ Mpc. The 
comparison with the power spectrum  of the galaxy distribution 
in the Stromlo-APM survey shows that at wavenumbers 
$k<0.06h$ Mpc$^{-1}$ the three-dimensional power spectrum 
computed by Lin et al. (1996) is probably underestimated.

As discussed by TE, the peak in the power spectrum of the
Stromlo-APM survey may be caused by the effects of 
uncertainty in the mean number density of galaxies
and may not reflect a real feature of the galaxy distribution. 
However, independent evidence for the presence of a 
preferred scale in the universe around $120h^{-1}$ Mpc comes from
an analysis of the distribution of galaxy clusters. Figure~2 shows 
the power spectrum of the distribution of galaxy clusters as 
determined by Einasto et al. (1997). The power spectrum is 
calculated for 869 Abell clusters with measured redshifts. The 
power spectrum of the distribution of galaxy clusters has a 
well-defined peak at the same wavenumber, $k_0=0.052h^{-1}$ Mpc, 
as the power spectrum of galaxies in the Stromlo-APM survey.
For wavenumbers $k>k_0$ the shape of the clusters power spectrum 
is similar to the shape of the power spectrum 
for galaxies in the Stromlo-APM survey. This comparison suggests 
that the peak observed in the power spectrum of the Stromlo-APM 
survey is a real feature in the distribution of galaxies.

\section{METHOD FOR DETERMINING THE VELOCITY POWER SPECTRUM
FROM REDSHIFT DATA}

To estimate the power spectrum of velocity fluctuations from a 
given power spectrum of galaxy clustering in redshift space we 
assume that: 1) the power spectrum of galaxy clustering 
in real space is $P_{gal} (k) = b^2 P(k)$, where $b$ is 
the bias factor; and 2) the relation between the power spectra of
density and velocity fluctuations is given by the linear theory 
relation (equation 9). We assume that these assumptions hold for 
wavenumbers $k<0.15h$ Mpc$^{-1}$ ($\lambda > 42h^{-1}$ Mpc) and 
examine the power spectrum of velocity fluctuations in this range. 

Galaxy peculiar velocities cause a distortion of the clustering 
pattern measured in redshift space compared to the true pattern 
in real space (see e.g. Kaiser 1987, Gramann, Cen, \& Bahcall
1993). To take account of the redshift-space distortions we use the 
following analytic model:
$$
P^s_{gal} (k) = (1 + 2\beta/3 + \beta^2/5) \,\, G^2(\beta,k\sigma_v) \,\, 
P_{gal}(k) ,
\eqno (14)
$$ 
where the parameter $\beta=f(\Omega_0)/b$ and the function 
$G$ is given by
$$
G^2(\beta,k\sigma_v)=
[{\sqrt\pi \, {\rm erf}(y) \over 8 y^5} \, (3\beta^2+4\beta y^2 +4 y^4) -  
$$
$$
-{\exp(-y^2) \over 4 y^4} \, (3 \beta^2 +2 \beta^2 y^2 +4 \beta y^2)]/
(1 + 2\beta/3 + \beta^2/5)\, ,
\eqno(15)
$$
where $y=k\sigma_v/H_0$. The first factor in equation (14) is expected
from linear theory (Kaiser 1987). The function $G(\beta,k\sigma_v)$ 
describes the suppression of the power spectrum on small scales as given 
by Peacock \& Dodds (1994). For $k \to 0$ (linear regime), 
the function $G(\beta,k\sigma_v) \to 1$. The small-scale peculiar 
velocities are assumed to be uncorrelated in position and are drawn 
from a Gaussian distribution with one-dimensional dispersion $\sigma_v$. 
Numerical simulations have shown that the analytic model (14) provides a 
good match to the peculiar velocity distortion in redshift space 
(Gramann, Cen, \& Bahcall 1993; TE). In the mixed dark matter model 
the redshift-space distortion can be fitted with approximation (14), 
using the parameter $\sigma_v \simeq 500$ km s$^{-1}$; in 
the low-density cold dark matter models we can describe the velocity
distortion in redshift-space using $\sigma_v$ in the range 
$200-350$ km s$^{-1}$,
depending on the amplitude of the power spectrum. The velocity dispersion, 
$\sigma_v$, depends on the power spectrum of density
and velocity fluctuations in the universe, but this relation is 
not linear. It can be determined using numerical simulations for a 
given model. In this paper we will examine the redshift-space distortion 
for various assumed values of $\sigma_v$.

Using approximation (14), the power spectrum of velocity fluctuations
is determined as
$$
V^2 (k) = {1 \over 2 \pi^2} \,\, H_0^2 \,\, F^2(\beta) \,\, 
G^{-2}(\beta,k\sigma_v)\,  k P_{gal}^s(k) , 
\, \, \, \, \, \,
F^2 (\beta) = {\beta^2 \over 1 + 2\beta/3 + \beta^2/5}.
\eqno (16)
$$
We use equation (16) to calculate the power spectrum of 
velocity fluctuations from a power spectrum of the galaxy 
distribution. 

The amplitude of the velocity fluctuations derived from the galaxy 
distribution depends on the parameter $\beta$. This parameter 
can be estimated on the basis of the observed mass function of  
galaxy clusters. The present data for the cluster mass function 
indicate that the parameter $\beta$ is lower than one, the preferred 
range being $\beta \simeq 0.4-0.7$ 
(e.g. Bahcall \& Cen 1993; White, Efstathiou, 
\& Frenk 1993; Eke, Cole, \& Frenk 1996). However, the parameter 
$\beta$ determined starting from the cluster mass function depends on 
the bias factor for galaxies on scales $r<10h^{-1}$ Mpc. This is not 
necessarily equal to the bias factor on the larger scales 
($k<0.15h^{-1}$ Mpc) examined in this paper. 

We have investigated the redshift-space distortion at the maximum 
wavenumber, $k=0.15h$ Mpc$^{-1}$, used to estimate the
power spectrum of velocity fluctuations. At this wavenumber, the
non-linear effect on the power spectrum in redshift space
can be quite large, especially if the galaxy velocity dispersion is high.
For a velocity dispersion 
$\sigma_v=600$ km s$^{-1}$, we find that the function $V(k)$ is 
enhanced by $\sim 17$\% compared to the linear theory prediction. 
For high velocity dispersions, the non-linear correction to 
the velocity spectrum at wavenumbers $k \le 0.15h$ Mpc$^{-1}$ is 
therefore important. However, the true one-dimensional rms velocity
of galaxies is probably significantly less than $600$ km s$^{-1}$.
This value of $\sigma_v$ corresponds to an rms pairwise velocity 
$\sigma_{12} \simeq 850-900$ km s$^{-1}$.
For a velocity dispersion $\sigma_v=400$ km s$^{-1}$, the function 
$V(k)$ is only $\sim 8$\% larger than expected from the linear approximation 
at a wavenumber $k=0.15h$ Mpc$^{-1}$. In this case we can use the 
linear approximation to estimate the power spectrum of velocity 
fluctuations at wavenumbers $k<0.15h$ Mpc$^{-1}$.

To determine the rms velocity of matter in the universe we use 
the following equation:
$$
v^2_{\rm rms} (r) = v^2_P(r)+v^2_L = 
\int V^2 (k) \exp(-r^2 \,k^2) \, {dk \over k} \, + v^2_L \,,
\eqno (17)
$$
where the function $v_P(r)$ describes the contribution of fluctuations 
derived from the galaxy power spectrum in a given redshift
survey, and the parameter $v_L$ describes the 
contribution from large-scale fluctuations in the universe which 
may exist on scales that are comparable to or greater than the
size of the redshift survey.  

To determine the function $v_P(r)$ in the Las Campanas survey
we use the velocity spectrum which is derived directly from the 
redshift data in the range $0.013 < k < 0.15h$ Mpc$^{-1}$ and on 
scales larger and smaller than this range we use an approximation (see
equation 18 below). 
Using this approximation, the contribution to the
velocity dispersion $v^2_{\rm rms}$, at a radius of $3h^{-1}$ Mpc, is 
$\sim 2.5$\% and $\sim 23$\% from fluctuations with wavenumbers $k < 0.013h$ Mpc$^{-1}$
and $k > 0.15h$ Mpc$^{-1}$ respectively. To determine the function 
$v_P(r)$ in the Stromlo-APM survey we use the velocity spectrum which 
is derived directly from the galaxy power spectrum in the range 
$0.007 < k < 0.15h^{-1}$ Mpc and outside this range we again use an 
approximation (see equation 22 below). Using this approximation, 
the contribution from 
fluctuations with wavenumbers $k<0.007h^{-1}$ Mpc is 
$\sim 2$\% and fluctuations at wavenumbers $k>0.15h^{-1}$ Mpc 
contribute $\sim 10$\% to the velocity dispersion at a radius 
$r=3h^{-1}$ Mpc.

Let us now investigate the power spectrum of velocity fluctuations
and the rms velocity of matter starting from the power spectrum 
of the galaxy distribution in the Las Campanas and Stromlo-APM 
redshift surveys.

\section{VELOCITY FLUCTUATIONS IN THE LAS CAMPANAS GALAXY SURVEY}

Figure~3a shows the function $V(k)$, computed from equation (16), 
for the power spectrum of galaxy clustering in the Las Campanas
redshift survey. The rms amplitude of velocity fluctuations is 
presented for the parameter $\beta=0.7$. To see the effect of 
redshift-space distortion in more detail, we investigated the 
function $V(k)$ for different values of velocity dispersion $\sigma_v$. 
Figure~3a shows the function $V(k)$ for three 
values of velocity dispersion: $\sigma_v=0, 400$, and $600$ km s$^{-1}$.
The power spectrum of velocity fluctuations derived from the
Las Campanas survey increases up to the wavenumber 
$k \simeq 0.06h$ Mpc$^{-1}$ ($\lambda \simeq 100h^{-1}$ Mpc) and 
flattens for larger wavenumbers. The maximum value for the function 
$V(k)$ is shifted to smaller scales in comparison with the power spectrum 
of the galaxy distribution and occurs in the range 
$0.08<k<0.1h$ Mpc$^{-1}$. To describe the power 
spectrum of velocity fluctuations derived from the Las Campanas 
survey we can use the fitting function 
$$
V^2(k)=2V^2 (k_0) \, (k/k_0)^{n+1} \,\, [1+ (k/k_0)^{n+m}]^{-1} \, ,
\eqno(18)
$$
where the parameters $k_0=0.06h$ Mpc$^{-1}$, $n=1.2$, $m=1.7$ and the 
value of the velocity power spectrum at a wavenumber $k_0$ is given by 
$$
V(k_0)=625 \, F(\beta) \, {\rm km \, s}^{-1} \, .
\eqno(19)
$$
For the parameter $\beta=0.7$ ($F(\beta)=0.56$), the rms amplitude 
of velocity fluctuations $V(k_0)=350$ km s$^{-1}$. 
At a wavenumber $k_0=0.06h$ Mpc$^{-1}$, the non-linear correction to 
the function $V(k)$ is small and can be neglected ($\sim 1.3$\% 
if $\sigma_v=400$ km s$^{-1}$ and $\sim 3$\% if $\sigma_v=600$ km s$^{-1}$). 
At smaller scales the function $V(k)$ is better fitted 
with index $m=1.7$ rather than the value  $m=1.8$ used in equation (13).
The fitting function (18) is consistent with the power spectrum in the 
Las Campanas survey at a confidence level of $>85\%$ 
(if $\sigma_v\leq 600$ km s$^{-1}$).

For comparison we show in Figure~3a the rms amplitude of velocity
fluctuations derived from peculiar velocities of galaxies. 
Kolatt \& Dekel (1997) compared the power spectra of density 
fluctuations derived from peculiar velocities with 
galaxy power spectra determined from various large galaxy
surveys and derived best-fitting values for the parameter $\beta$ in
the range $0.77 - 1.21$. The power 
spectrum of the galaxy distribution in the Las Campanas redshift
survey is consistent with the power spectrum estimated from peculiar 
velocities when the parameter $\beta \approx 1.0$.

Let us now consider the rms peculiar velocity of matter, starting from the
power spectrum of the galaxy distribution. Figure~3b shows the rms 
velocity computed from equation (17) for the velocity spectra presented 
in Figure~3a, assuming that the parameter $v_L=0$. The fluctuations at 
wavenumbers $k<k_0$ contribute $\sim 33\%$ to velocity dispersion of 
galaxy systems and $\sim 67$\% of the velocity dispersion is generated
on smaller scales. The rms peculiar velocity of matter at the smoothing 
radius $r=3h^{-1}$ Mpc can be written as
$$
v_{\rm rms} (r=3h^{-1} {\rm Mpc}) = (870 \pm 90) \,\, G_{\rm int} (\sigma_v) 
\,\, F(\beta) \,\, {\rm km \, s}^{-1} \,.
\eqno (20)
$$
The function $G_{\rm int}(\sigma_v)$ describes the correction due to the non-linear redshift-space 
distortions. For the velocity dispersion $\sigma_v=400$ km s$^{-1}$ 
and $\sigma_v=600$ km s$^{-1}$, the function $G_{\rm int}=1.02$ and 
$1.05$, respectively. (The non-linear correction also depends
on the parameter $\beta$, but this dependence is very weak and 
is neglected here). For the parameters $\beta=0.7$
and $\sigma_v=400$ km s$^{-1}$, the rms peculiar velocity 
$v_{\rm rms}(r=3h^{-1} {\rm Mpc}) = (498 \pm 50)$ km s$^{-1}$.
The small-scale velocity dispersion of the matter is 
consistent with the observed dispersion of galaxy clusters when the 
parameter $\beta$ is in the range $0.6-0.7$.

Figure~3c shows the bulk velocities that correspond to 
the power spectrum of the galaxy distribution in the Las Campanas survey 
for the parameter $\beta=0.7$. The bulk velocities were computed using 
equation (10). The bulk velocity at a radius $r=60h^{-1}$ Mpc 
can be written as
$$
v_b (r=60h^{-1} {\rm Mpc}) = (305 \pm 75) \,\, F(\beta) \,\, 
{\rm km \, s}^{-1} .
\eqno (21)
$$
For the parameter $\beta=0.7$, the rms amplitude of the bulk flow 
averaged on a scale $r=60h^{-1}$ Mpc is 
$(170 \pm 40)\, {\rm km \, s}^{-1}$. 

For comparison, we plot in Figure~3c the observed bulk velocities 
derived from the Mark III catalog of peculiar velocities 
for radii $30$, $40$, $50$ and $60h^{-1}$ Mpc (Dekel 1994). The 
observed velocities are determined in a sphere centered on the Local 
Group and represent a single measurement of the bulk flow on large 
scales. The average velocity of galaxies in the sphere of radius 
$r=60h^{-1}$ Mpc around us is estimated as $370 \pm 80$ km s$^{-1}$.
Assuming the distribution of bulk velocities is a Maxwellian 
distribution with rms velocity $\simeq 170$ km s$^{-1}$, the probability 
of measuring a bulk velocity $\ge 300$ km s$^{-1}$ is only 2.5\%.

The difference between the small-scale velocity dispersion of galaxy 
systems and the large-scale velocity dispersion at radius 
$r=60h^{-1}$ Mpc is determined by the amplitude of velocity 
fluctuations at intermediate wavenumbers $k\sim 0.05-0.1$ 
($\lambda \sim 120-60h^{-1}$ Mpc). If the rms velocity of galaxy 
systems is $\sim 500$ km s$^{-1}$ and the rms amplitude of the 
bulk flow, averaged over a scale of $r=60h^{-1}$ Mpc, is 
$\ge 300$ km s$^{-1}$, then the contribution from velocity 
fluctuations at intermediate wavenumbers must be 
$\le 400$ km s$^{-1}$. This situation is consistent with the 
amplitude of velocity fluctuations derived from the Las Campanas 
survey, only when the parameters $\beta \le 0.6$ and $v_L \ge 0$.
If the amplitude of the large-scale velocity fluctuations at
wavenumbers $k<0.06h$ Mpc$^{-1}$ is higher than that estimated starting from
the power spectrum of the galaxy distribution in the Las Campanas survey, 
the observed rms peculiar velocity of clusters is consistent with a lower 
amplitude for the velocity fluctuations at smaller wavelengths and with a
lower value of the parameter $\beta$.

\section{VELOCITY FLUCTUATIONS IN THE STROMLO-APM GALAXY SURVEY}

Figure~4a shows the function $V(k)$, computed from equation (16) 
for the power spectrum of galaxy clustering in the Stromlo-APM 
redshift survey. The rms amplitude of velocity fluctuations is presented
for the parameter $\beta=0.55$. As for the Las Campanas survey, we
use the parameter $\beta$ which is consistent with the 
observed dispersion of galaxy clusters (see below). Figure~4a shows the
results for velocity dispersions $\sigma_v=0, 400$, and 
$600$ km s$^{-1}$.

The power spectrum of velocity fluctuations, like the power 
spectrum of the galaxy distribution in the Stromlo-APM redshift 
survey, peaks at a wavenumber $k_0=0.052h$ Mpc$^{-1}$ (or at 
a wavelength $\lambda = 120h^{-1}$ Mpc). At smaller wavelengths 
the function $V(k)$ declines, reaching a minimum value at 
$k=0.127h$ Mpc$^{-1}$. To describe the power spectrum of velocity 
fluctuations in the Stromlo-APM survey, we use the function
$$
V^2(k)=\cases {V^2(k_0)(k/k_0)^{n+1}, &if $k<k_0$;\cr  
V^2(k_0) (k/k_0)^{m+1}, &if $k > k_0$ , \cr}
\eqno(22)
$$
where the parameters $n=0.5$ and $m=-2$ as in equation (12) and 
the value of the velocity spectrum at its maximum is given by
$$
V(k_0)=915 \, F(\beta) \, {\rm km \, s}^{-1} \, .
\eqno(23)
$$
For the parameter $\beta=0.55$ ($F(\beta)=0.46$) the maximum value 
for the velocity rms amplitude is $V(k_0)=420$ km s$^{-1}$.
The function (22) is consistent with the power spectrum 
in the Stromlo-APM survey at a confidence level of $\geq 70$\% 
(assuming that $\sigma_v\leq 600$ km s$^{-1}$).

The power spectrum of the galaxy distribution in the Stromlo-APM redshift
survey is consistent with the power spectrum estimated from peculiar 
velocities of galaxies by Kolatt \& Dekel (1996) when the parameter 
$\beta \approx 0.8-0.9$. For the parameter $\beta=0.55$, 
these two power spectra are only consistent 
at wavenumbers $k\sim 0.06h$ Mpc$^{-1}$. At smaller
scales the function $V(k)$ derived from the distribution of galaxies 
is smaller (by a factor of $\sim 1.6$ at $k=0.1h$ Mpc$^{-1}$). 

Figure~4b shows the rms velocity computed from equation (17) for the 
velocity spectra presented in Figure~4a, if the parameter 
$v_L=0$. By substituting approximation 
(22) into (17) we find that for the index $m=-2$, the velocity
dispersion 
$$
v^2_{\rm rms} (r) = V^2 (k_0) [ {1 \over n+1} + g(rk_0)] \, , \,\,\,\,\,
g(rk_0)=\exp (-r^2 k_0^2) - \sqrt\pi\, r k_0 [1 - {\rm erf}(rk_0)] \, .
\eqno (24)
$$
Figure~4b shows that equation (24) provides a good match to the velocity 
dispersion in the Stromlo-APM survey. The first factor in equation (24) 
gives the contribution from large-scale velocity fluctuations at 
wavenumbers $k<k_0$.
To compute this, we assumed that $\exp(-r^2 k^2) = 1$ for $k<k_0$. 
For an index $n=0.5$, the contribution of large-scale 
fluctuations is $\sim 1/(n+1) = 2/3$. The function $g(rk_0)$ gives the 
contribution from small-scale velocity fluctuations at wavenumbers $k>k_0$.
For the parameters $k_0=0.052h$ Mpc$^{-1}$ and $r=3h^{-1}$ Mpc, the 
function $g(rk_0) \approx 3/4$. Therefore, the 
large-scale fluctuations at wavenumbers $k<k_0$ and small-scale 
fluctuations at wavenumbers $k>k_0$ give similar contributions 
(47\% and 53\%, respectively) to the velocity dispersion of matter 
smoothed at radius $r=3h^{-1}$ Mpc and the rms velocity of matter
$$
v^2_{\rm rms} (r=3h^{-1} {\rm Mpc}) \approx 1.4 \, V^2 (k_0) \, . 
\eqno (25)
$$
The fluctuations at wavenumbers $0.04<k<0.07h$ Mpc$^{-1}$, 
around the maximum at $k_0=0.052h$ Mpc$^{-1}$, contribute $\sim 33\%$ 
to the velocity dispersion of galaxy systems.

For the Stromlo-APM survey, the rms peculiar velocity 
of matter at radius $r=3h^{-1}$ Mpc depends on the parameter $\beta$ as
$$
v_{\rm rms} (r=3h^{-1} {\rm Mpc}) = (1080 \pm 160) \,\, G_{\rm int} (\sigma_v) 
\,\, F(\beta) \,\, {\rm km \, s}^{-1} .
\eqno (26)
$$
Since the contribution from small-scale fluctuations in the Stromlo-APM
survey is less important than in the Las Campanas survey, 
the function $G_{\rm int} (\sigma_v)$ for the Stromlo-APM survey is 
also somewhat smaller. For velocity dispersions 
$\sigma_v=400$ km s$^{-1}$ and $\sigma_v=600$ km s$^{-1}$,
the function $G_{\rm int}=1.015$ and $1.035$, respectively. 
For the parameter $\beta=0.55$ and $\sigma_v=400$ km s$^{-1}$, the rms 
peculiar velocity $v_{\rm rms}(r=3h^{-1} {\rm Mpc}) = (505 \pm 75)$ km s$^{-1}$.
The small-scale velocity dispersion of matter is consistent with the 
observed dispersion of galaxy clusters when the parameter $\beta$ is in
the range $0.5-0.6$.

Figure~4c shows the bulk velocities that correspond to the power
spectrum of the galaxy distribution in the Stromlo-APM survey 
for the parameter $\beta=0.55$. The bulk velocity at radius 
$r=60h^{-1}$ Mpc can be written in the form
$$
v_b (r=60h^{-1} {\rm Mpc}) = (535 \pm 145) \,\, F(\beta) \,\, 
{\rm km \, s}^{-1} .
\eqno (27)
$$
For the parameter $\beta=0.55$, the bulk velocity 
$v_b (r=60h^{-1} {\rm Mpc}) = (245 \pm 70)\, {\rm km \, s}^{-1}$. 
For an rms velocity $\simeq 250$ km s$^{-1}$, the probability
of measuring a bulk velocity larger than $300$ km s$^{-1}$ 
is about 23\%. The probability of measuring a velocity larger than
$350$ km s$^{-1}$ is 12\%. 

Until this point we have assumed that the power spectrum of the galaxy
distribution decreases monotonically for wavelengths $\lambda
>120h^{-1}$ Mpc. There may, however, be a significant contribution to the 
power from fluctuations on scales comparable to, or greater than, the size
of the Stromlo-APM survey. In this case the observed rms cluster
peculiar velocity is consistent with a smaller amplitude for the
velocity fluctuations at smaller wavelengths and thus, with a lower
value of the parameter $\beta$. 

Figure~5 shows the properties of the peculiar velocity field for the 
parameters $\beta=0.45$ and $v_L\ge0$.
For the parameter $\beta=0.45$ ($F(\beta)=0.39$), the maximum value for
the velocity rms amplitude is $V(k_0)=350$ km s$^{-1}$. The rms 
velocity of matter at the smoothing radius $r=3h^{-1}$ Mpc is 
$v_{\rm rms} \simeq 420$ km s$^{-1}$, if the parameter $v_L=0$ and
$v_{\rm rms}(r=3h^{-1} {\rm Mpc})\simeq 500$ km s$^{-1}$,
if there is an additional contribution from the large-scale 
fluctuations in the universe characterized by
$v_L=270$ km s$^{-1}$. For the parameter $\beta=0.4$, we obtain a 
similar estimate for the velocity dispersion of galaxy systems ($500$
km s$^{-1}$), by assuming the value of $v_L \simeq 330$ km s$^{-1}$. 
In the latter case the fluctuations determined from the galaxy 
distribution in the Stromlo-APM redshift survey contribute only 
$\sim 58$\% to the velocity dispersion of galaxy systems and the rest 
comes from scales greater than the size of the redshift survey.

Figure~5c shows the bulk velocities that correspond to the power
spectrum of the galaxy distribution in the Stromlo-APM survey for the
parameter $\beta=0.45$. The bulk velocity at a radius $r=60h^{-1}$ Mpc
is $\simeq 210$ km s$^{-1}$, if the parameter $v_L=0$,
and the rms amplitude of the bulk flow increases to 
$\simeq 340$ km s$^{-1}$, if the parameter $v_L=270$ km s$^{-1}$. 
(Here we assumed that the contribution from large-scale fluctuations 
is similar at radii $r=3h^{-1}$ Mpc and $r=60h^{-1}$ Mpc. The 
contribution at larger radii can be somewhat smaller, depending 
how the large-scale power on wavenumbers $\lambda \ge 120h^{-1}$ Mpc 
is distributed.)

\section{DISCUSSION AND SUMMARY}

In this paper we have examined the power spectrum of velocity fluctuations
in the universe starting from the peculiar velocities of galaxies and
clusters of galaxies, and from the power spectrum of the galaxy 
distribution in redshift surveys.
There are various ways of interpreting the data:
 
(1) The power spectrum of velocity fluctuations follows a power law,
$V^2(k) \sim k^2$, on 
large scales, achieves a maximum at wavenumbers $k_0 \sim
0.05-0.06h$ Mpc$^{-1}$, and declines as a power law,  $V^2(k) \propto k^{-0.8}$,
on smaller scales. The value of the function $V(k)$ at its
maximum is $\sim 500$ km s$^{-1}$, and the rms velocity of matter 
smoothed with a Gaussian of radius $3h^{-1}$ Mpc is $\sim 700$ km s$^{-1}$. 
This power spectrum of velocity fluctuations is consistent with the 
power spectrum of density fluctuations derived by Kolatt \& Dekel (1997) 
from peculiar velocities of galaxies, and with the power spectrum of 
the galaxy distribution in redshift surveys provided the parameter 
$\beta$ is in the range $0.8-1.0$. Data for peculiar velocities of 
galaxies yield the rms amplitude of velocity fluctuations 
$V(k)=414\pm 52$ km s$^{-1}$ at the wavenumber $k=0.17h$ Mpc$^{-1}$ 
and the velocity rms amplitude increases to 
$V(k)=502\pm 96$ km s$^{-1}$ at $k=0.06h$ Mpc$^{-1}$ (see Figure~1).

This power spectrum of velocity fluctuations is predicted in 
a mixed cold+hot dark matter model (CHDM) with density parameter 
$\Omega_0=1.0$. Figure~6a shows the rms amplitude of velocity 
fluctuations predicted in CHDM models with neutrino densities 
$\Omega_{\nu}=0.2$ and $\Omega_{\nu}=0.3$.
We have used the transfer function computed by Pogosyan \& Starobinsky (1995)
and the COBE normalization derived by Bunn \& White (1997).
The initial fluctuations are assumed to be adiabatic and scale-invariant 
with $n=1$. The baryonic density $\Omega_B=0.05$ and $h=0.5$. 
The power spectrum of velocity fluctuations predicted in these models 
is in good agreement with fluctuations derived from peculiar 
velocities of galaxies. This model is not consistent with the observed 
rms peculiar velocity of galaxy clusters determined by Bahcall \& Oh (1996).

(2) The shape of the power spectrum of velocity fluctuations is similar
to that in model (1), but the amplitude of the power spectrum is lower.
The transition between positive and negative spectral indices is smooth,
without the peak at wavelength $\lambda \sim 120h^{-1}$ Mpc.
The rms velocity of matter on scales $\sim 3h^{-1}$ Mpc is in the range
$450-500$ km s$^{-1}$. This rms velocity is consistent with the power
spectrum of the galaxy distribution in the Las Campanas redshift survey
when the parameter $\beta$ is in the range $0.6-0.7$. The rms amplitude
of velocity fluctuations is $\simeq 350$ km s$^{-1}$ at a wavelength
$\lambda \simeq 100h^{-1}$ Mpc and the rms amplitude of the bulk flow
on a scale of $\sim 60h^{-1}$ Mpc is $\simeq 170$ km s$^{-1}$. 
This value is not consistent with the observed bulk velocity of 
galaxies.

A smooth power spectrum of velocity fluctuations is predicted in
low-density cold dark matter (CDM) models. 
Figure~6b shows the rms amplitude of velocity fluctuations predicted 
in a flat CDM model with density parameter $\Omega_0=0.3$, baryonic 
density $\Omega_B=0.0125h^{-2}$ and a normalized Hubble constant 
$h=0.65$. In this model the rms peculiar velocity of matter
$v_{\rm rms} (r=3h^{-1}{\rm Mpc}) =480$ km s$^{-1}$ and
the bulk velocity $v_{b}(r=60h^{-1} {\rm Mpc})= 265$ km s$^{-1}$.
For comparison, we show in Figure~6b the function $V(k)$ 
derived from the Las Campanas and Stromlo-APM redshift surveys for the
parameter $\beta=0.5$. In the $\Omega_0=0.3$ model, this value of
$\beta$ gives a bias parameter $b\approx 1.0$. The amplitude of
velocity fluctuations predicted in the low-density CDM model is 
consistent with the power spectrum of the galaxy distribution 
for wavenumbers $k>0.06h^{-1}$ Mpc. This model is not consistent with
the power spectrum of density fluctuations derived by Kolatt \& Dekel
(1997) from peculiar velocities of galaxies. 

(3) There is a peak in the power spectrum of velocity fluctuations in 
the universe at a wavelength $\lambda_0 \simeq 120h^{-1}$ Mpc 
(or at a wavenumber $k_0\simeq 0.05h$ Mpc$^{-1}$) and on larger scales 
the power spectrum decreases with an index $n\simeq 0.5-1.0$. The maximum 
value of the function $V(k)$ is $\sim 420$ km s$^{-1}$ at a wavelength 
$\lambda=120h^{-1}$ Mpc. The bulk velocity in this model is 
$v_{b}(r=60h^{-1} {\rm Mpc})\simeq 250$ km s$^{-1}$.
The power spectrum of density fluctuations derived from peculiar velocities 
of galaxies by Kolatt \& Dekel (1997) is correct on large scales 
$\lambda \sim 100h^{-1}$Mpc, but overestimated on smaller scales. 

This power spectrum of velocity fluctuations is 
consistent with the power spectrum of the galaxy distribution in the 
Stromlo-APM redshift survey provided the parameter $\beta$ is in the range 
$0.5-0.6$ (see Figure 4). If the bias parameter $b\approx 1.0$, this value of 
$\beta$ corresponds to a density parameter 
$\Omega_0 \approx 0.4$. This power spectrum of velocity fluctuations is also 
consistent with the observed rms peculiar velocity of galaxy clusters.
The small-scale fluctuations at wavelengths $\lambda<\lambda_0$ and 
large-scale fluctuations at wavelengths $\lambda>\lambda_0$ give
similar contributions to the velocity dispersion of galaxy systems.

The power spectrum of density and velocity fluctuations in the universe
depends on the physical processes in the early universe. 
The peak in the power spectrum of the galaxy distribution at
wavelength $\lambda \simeq 120h^{-1}$ Mpc may be generated during the
era of radiation domination or earlier. One possible explanation for
the presence of such a peak in the power spectrum is an inflationary 
scenario with a scalar field whose potential has a localized feature
around some value of the field (Starobinsky 1992). In this scenario,
the value of the corresponding characteristic scale in the universe is
a free parameter, but the form of the power spectrum around this
scale serves as a discriminating characteristic.

(4) There is a peak in the power spectrum of velocity fluctuations in 
the universe at the wavelengths $\lambda \simeq 120h^{-1}$ Mpc as in
model (3), but on larger scales the amplitude of fluctuations is 
higher than that estimated starting from the power spectrum of the galaxy 
distribution in the Stromlo-APM redshift survey and approximation (22). 
For example, there could be another peak in the power spectrum of velocity 
fluctuations at wavelengths $\lambda >200h^{-1}$ Mpc. In this 
case the fluctuations on large scales contribute significantly 
to the velocity dispersion of galaxy 
systems. The observed rms cluster peculiar velocity is consistent 
with a smaller amplitude for the velocity fluctuations at 
intermediate wavelengths $\lambda \sim 60-120h^{-1}$ Mpc and thus, 
with a lower value of the parameter $\beta$. For the parameter 
$\beta$ in the range $0.4-0.5$, the observed rms cluster peculiar 
velocity is consistent with the rms amplitude of the 
bulk flow $\simeq 340$ km s$^{-1}$ at the scale $60h^{-1}$ Mpc. 
In this case the value of the function $V(k)$ at wavelength 
$\lambda=120h^{-1}$ Mpc is $\simeq 350$ km s$^{-1}$.
The power spectrum of velocity fluctuations in this model is 
not consistent with the power spectrum derived from peculiar velocities 
of galaxies.

Available data are insufficient to rule out any of the possibilities 
listed here. Direct measurements of the density parameter 
indicate that the mean density in the universe is lower than critical, 
the preferred range being $\Omega_0 \simeq 0.3-0.5$
(e.g. Dekel, Burstein \& White 1996). 
If the density parameter $\Omega_0 \simeq 0.4$, we can exclude 
the first model. This model predicts that the clusters of galaxies move
with high peculiar velocities and the rms velocity of clusters is 
$\sim 750$ km s$^{-1}$. Accurate peculiar velocities of galaxy clusters 
can serve as a discriminating test for this model. Larger redshift 
surveys, such as the Sloan Digital Survey (Gunn \& Weinberg 1995),
are required to accurately determine the power spectrum of the galaxy distribution 
on scales $\lambda >100h^{-1}$ Mpc and so
distinguish between the models listed here.

\acknowledgements

I thank Avishai Dekel, Idit Zehavi, Helen Tadros, Huan Lin and Michael
Vogeley for providing the power spectrum data, as well as Neta Bahcall, 
Jaan Einasto, Enn Saar and Helen Tadros for useful discussions. This work 
has been supported by the ESF grant 97-2645.

\newpage
\begin{center}
FIGURE CAPTIONS
\end{center}

\figcaption[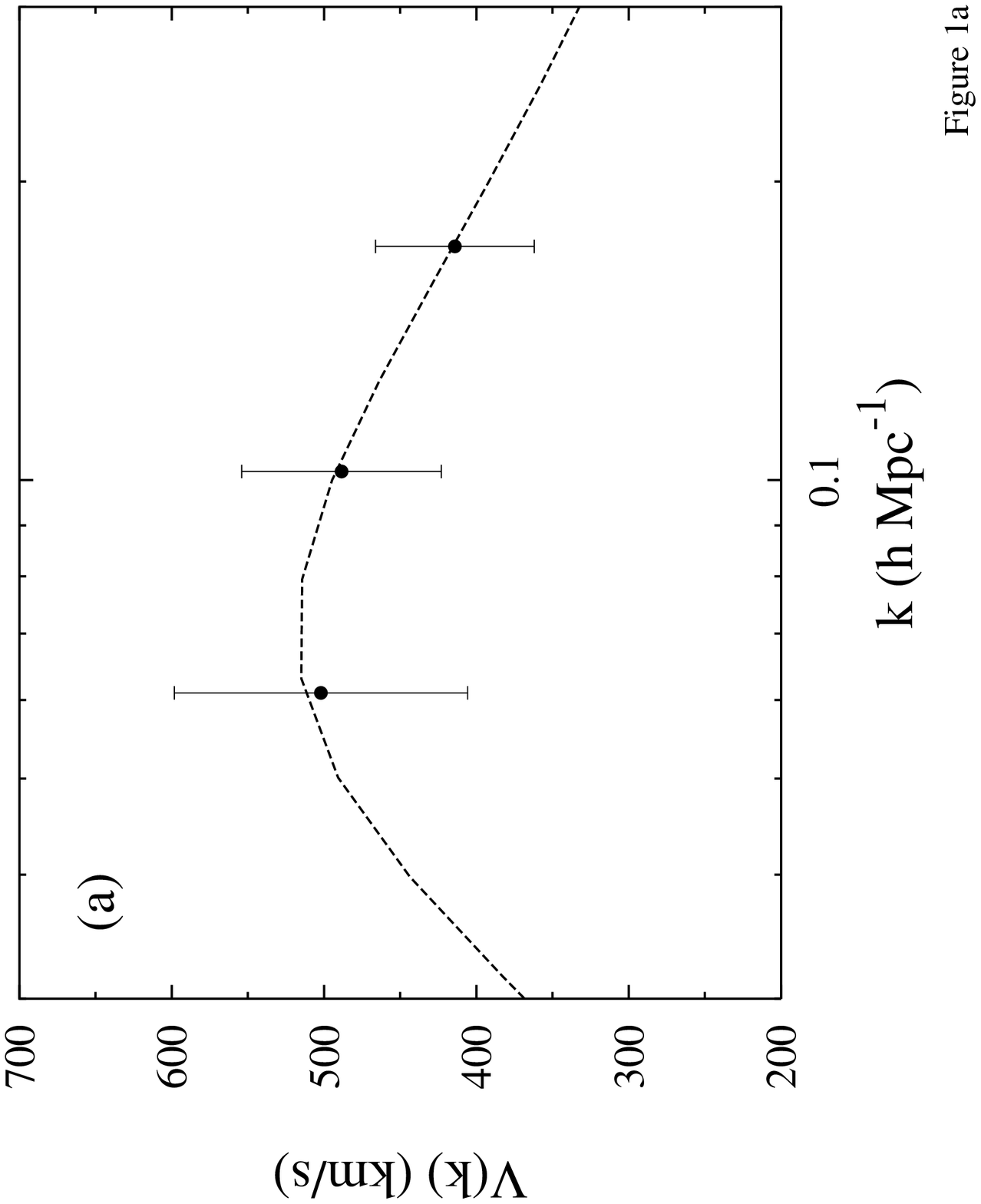,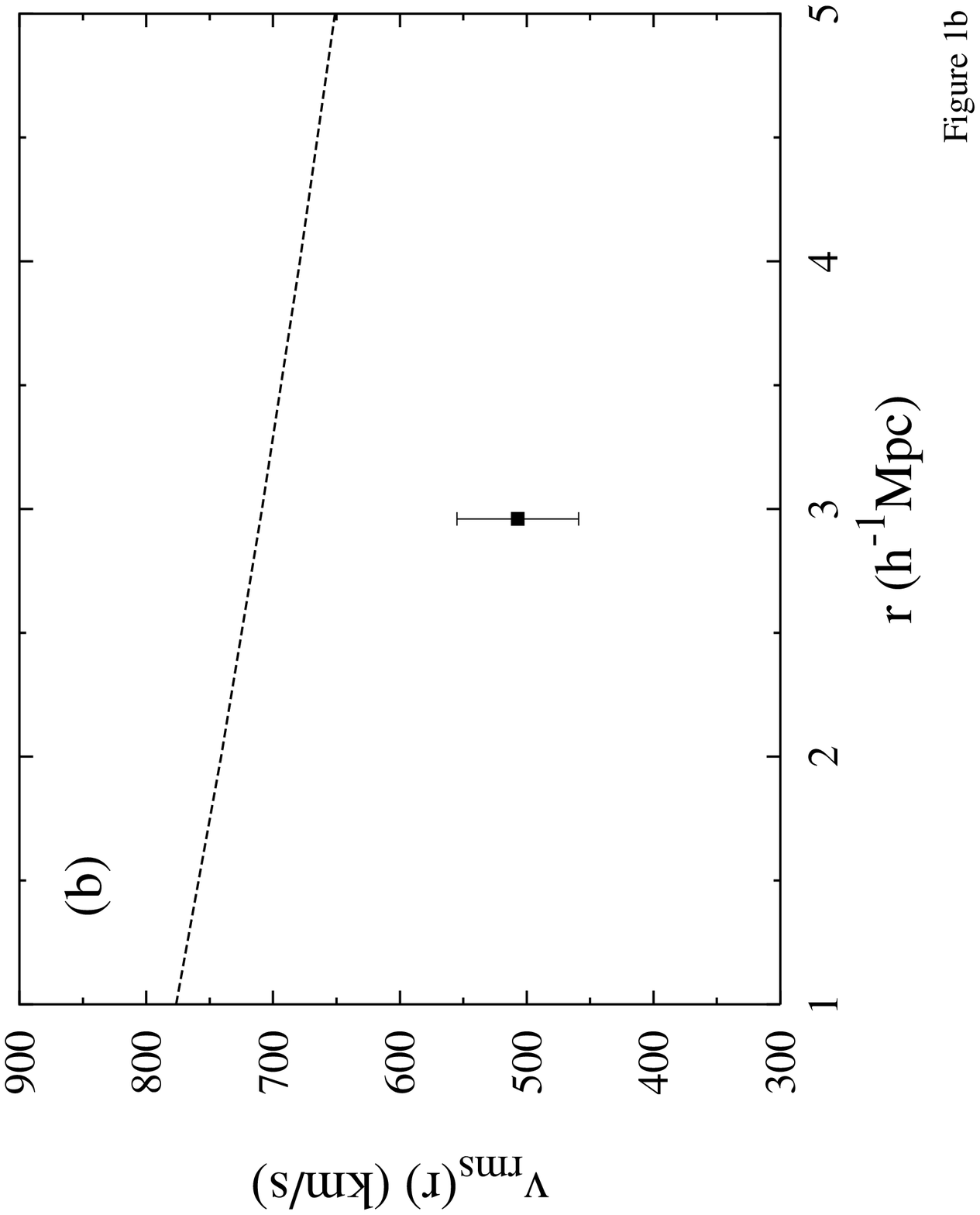]{(a) The rms amplitude of velocity 
fluctuations derived 
from peculiar velocities of galaxies (filled circles).
The dashed line is the fitting function (11). (b) The rms peculiar
velocity of matter for the velocity spectrum (11) (dashed line).
The filled square shows the observed rms peculiar velocity of
galaxy clusters.}

\figcaption[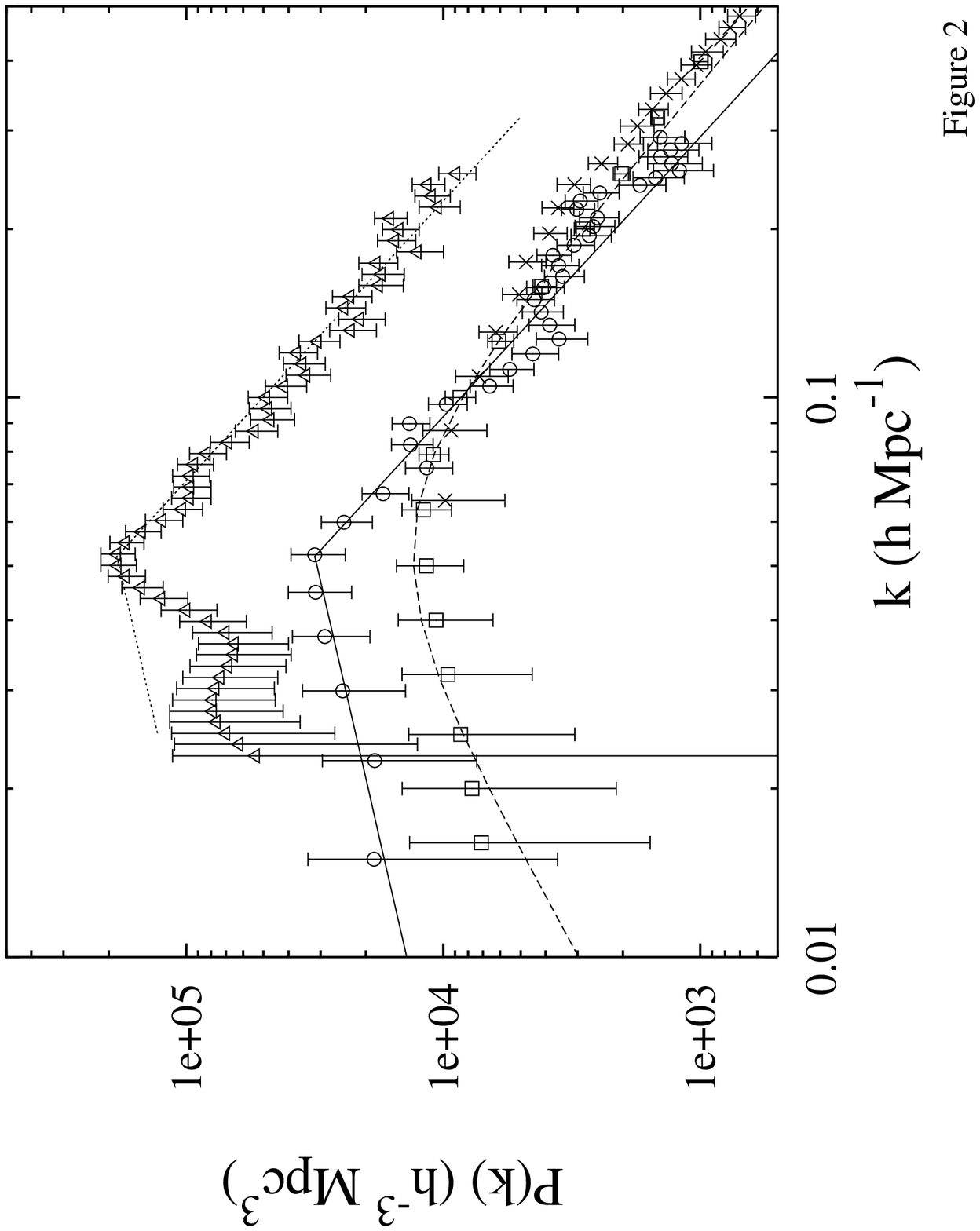]{The power spectrum of the galaxy distribution. 
Open circles and 
squares show the power spectrum of the galaxy distribution in
the Stromlo-APM and Las Campanas redshift survey, respectively.
To describe the power spectrum in these surveys we use the
functions (12) (solid line) and (13) (dashed line). Crosses show
the power spectrum of the galaxy distribution in the SSRS2+CfA2
redshift survey. For comparison, the open triangles describe the power
spectrum of the distribution of galaxy clusters. The dotted line is the
function (12) multiplied by the factor of $\sim 6$.}

\figcaption[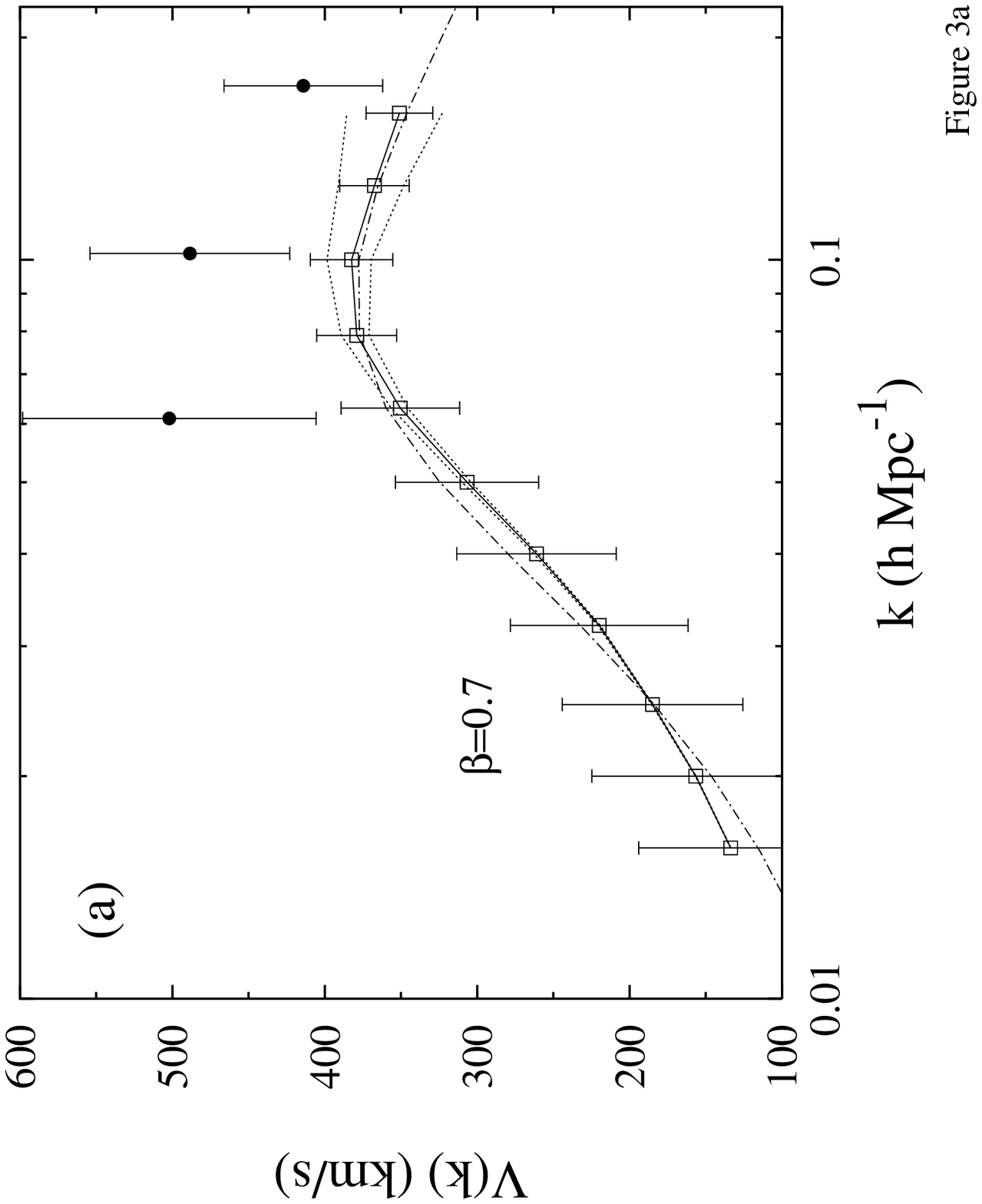,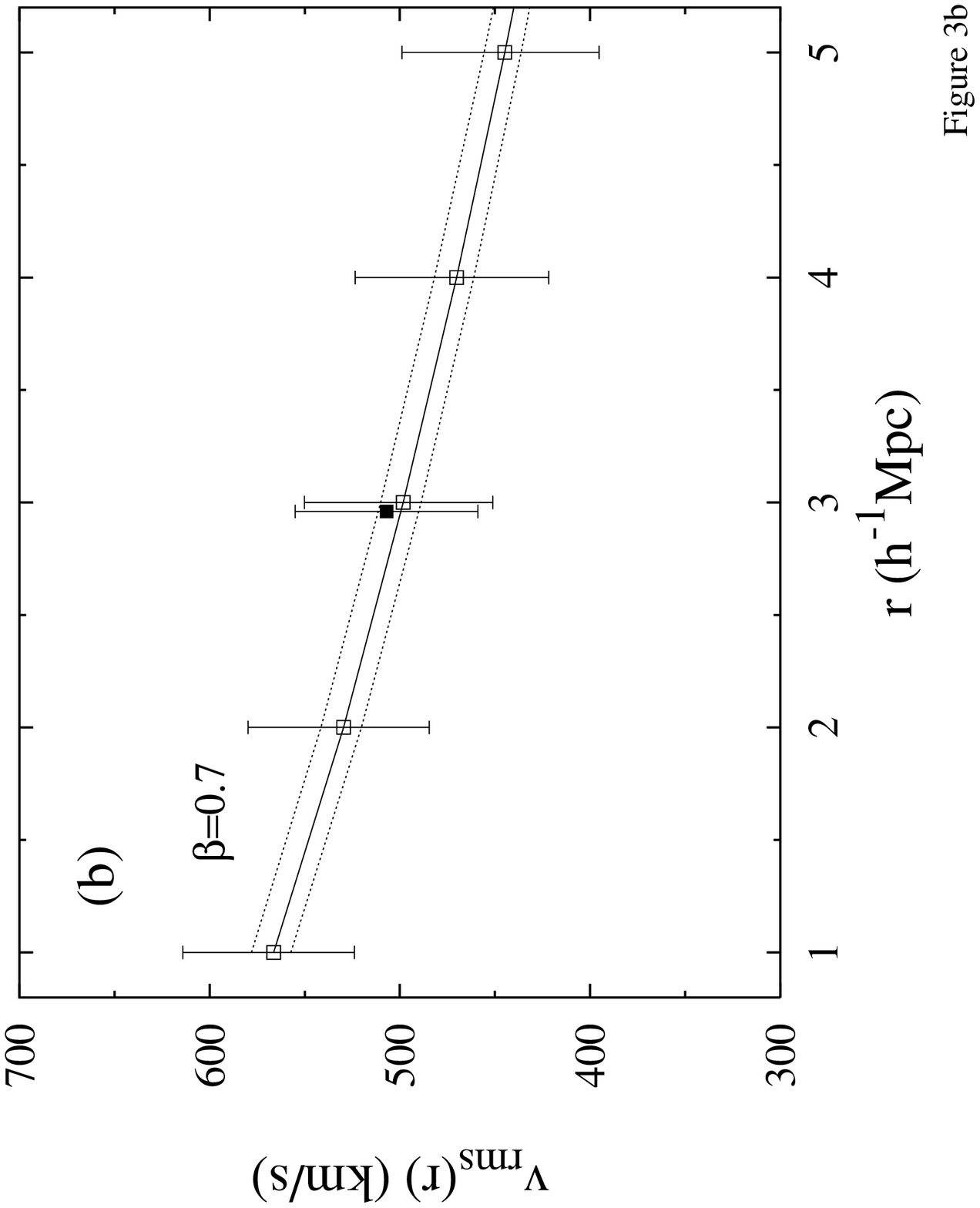,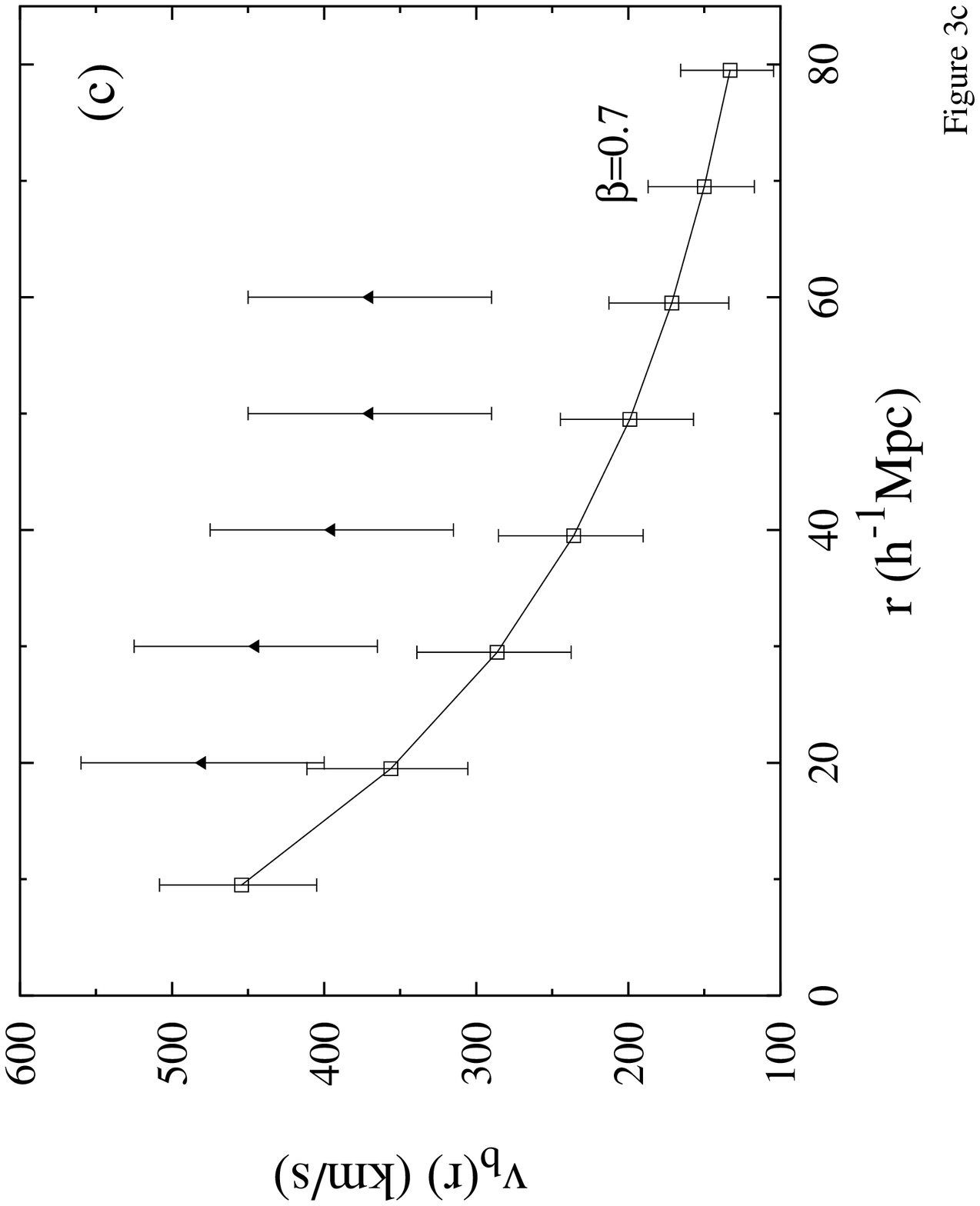]{Velocity fluctuations 
in the Las Campanas redshift survey
for the parameters $\beta=0.7$ and $\sigma_v=400$ km s$^{-1}$ (open
squares with solid line in each panel). (a) The rms amplitude of velocity 
fluctuations. Dotted lines show the function $V(k)$ for the velocity 
dispersion $\sigma_v=0$ (lower curve) and $\sigma_v=600$ km s$^{-1}$ 
(upper curve). The dot-dashed line describes the approximation (18). Filled 
circles demonstrate the velocity rms amplitude derived from peculiar 
velocities. (b) The rms peculiar velocity of matter for the 
velocity spectra presented in panel (a). Filled square demonstrates the 
observed rms cluster peculiar velocity. (c) The rms amplitude of the 
bulk flow. Filled triangles demonstrate the observed bulk velocities 
derived from peculiar velocities of galaxies.}

\figcaption[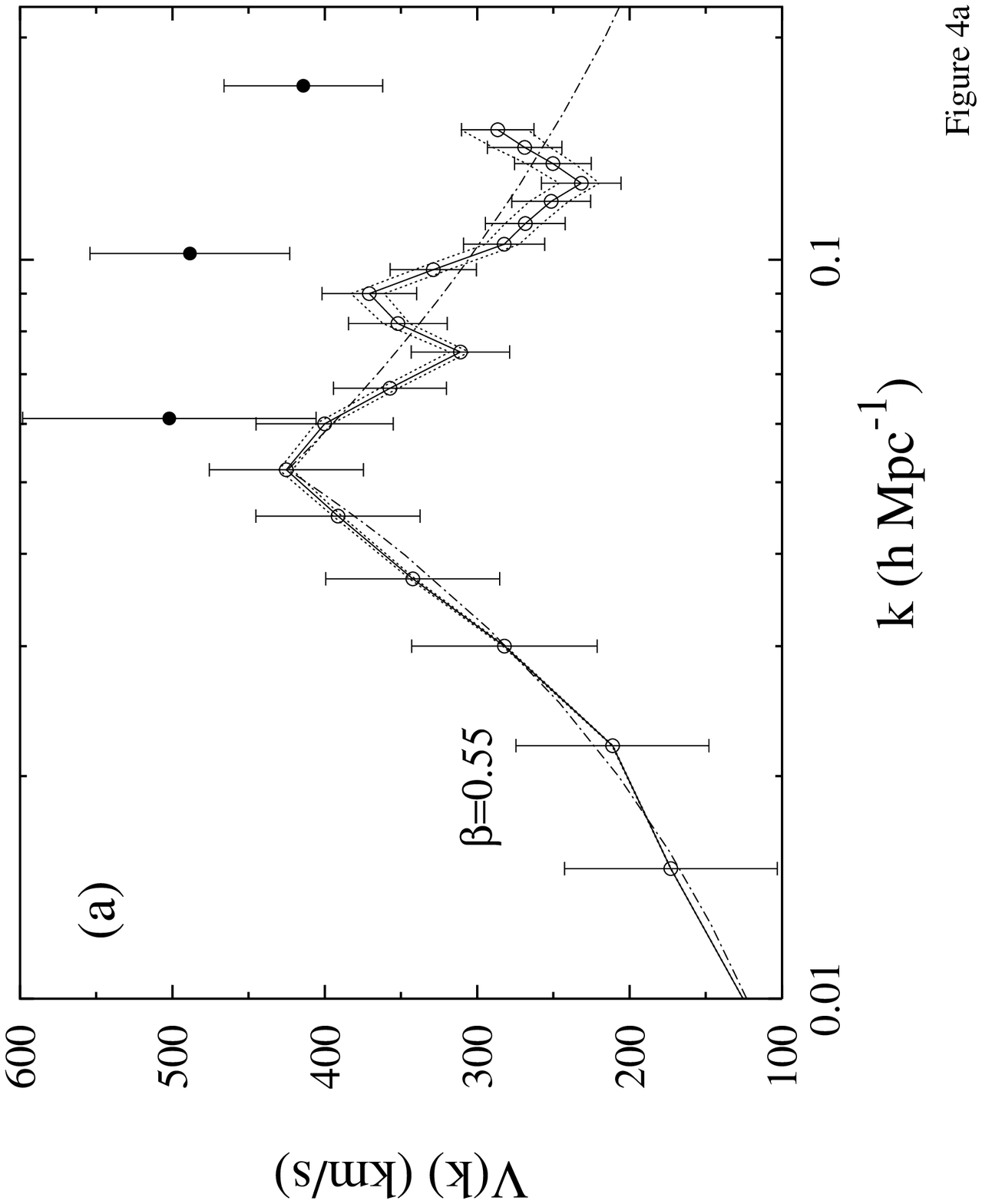,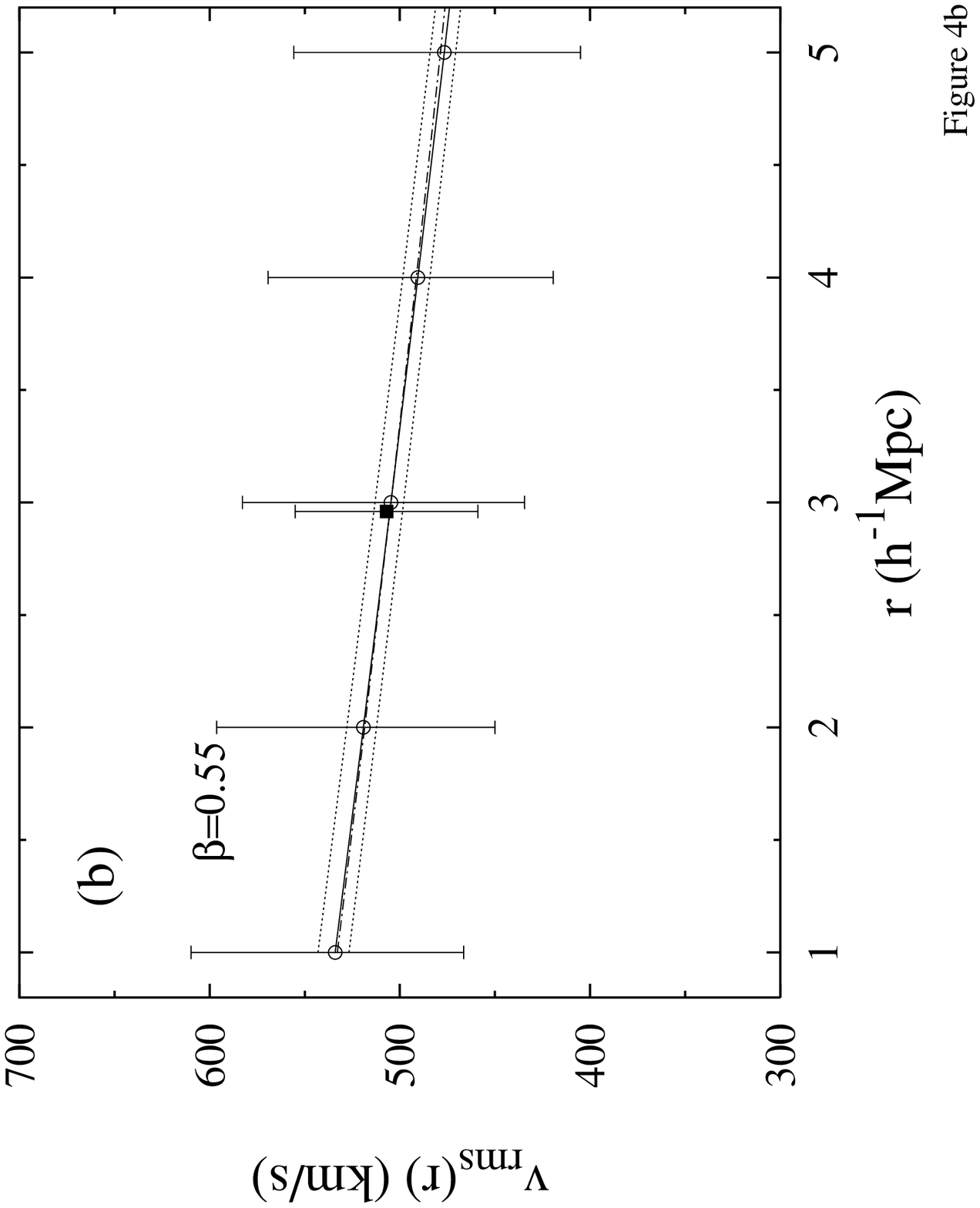,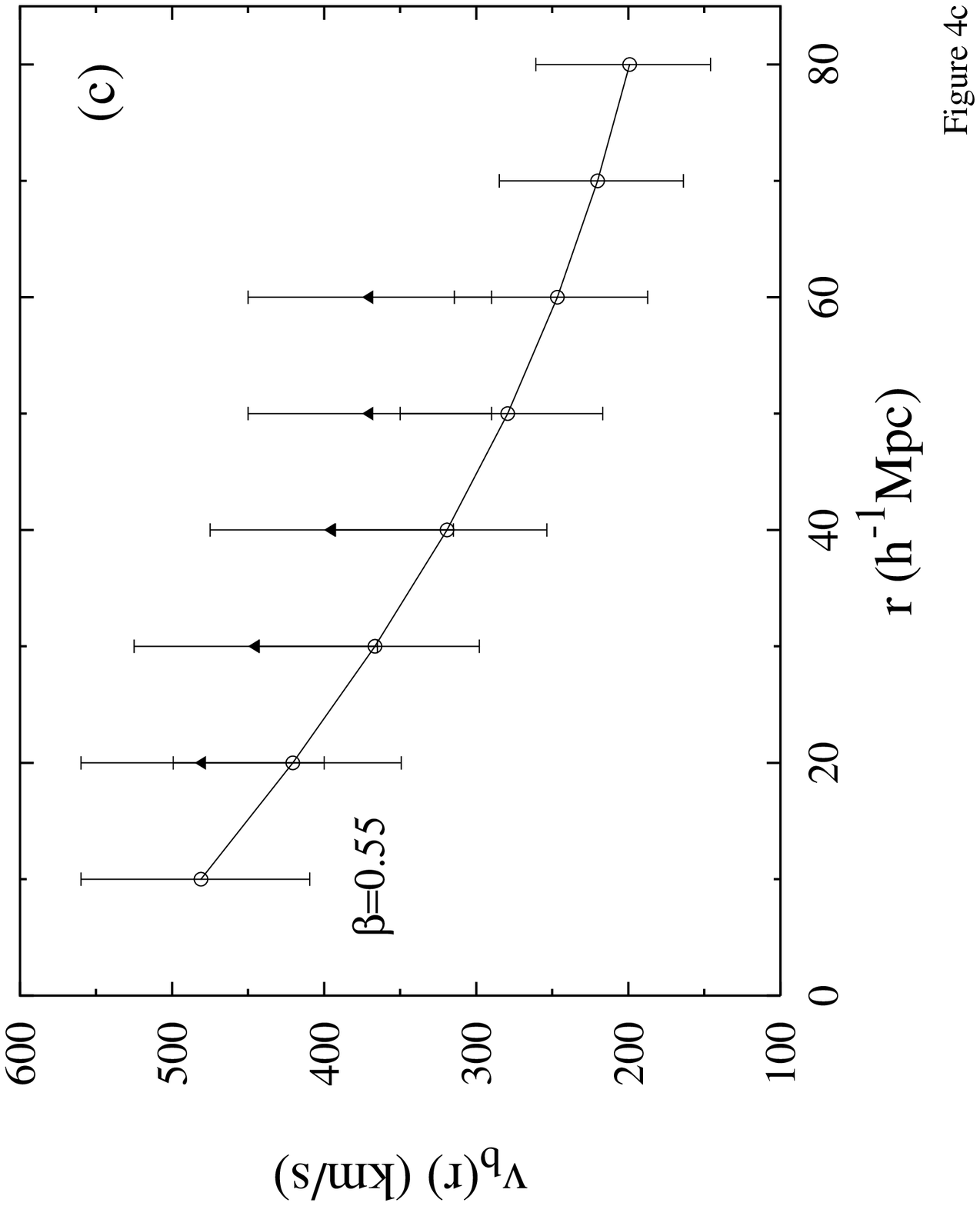]{Velocity fluctuations in the 
Stromlo-APM redshift survey
for the parameters $\beta=0.55$ and $\sigma_v=400$ km s$^{-1}$ (open
circles with solid line in each panel). (a) The rms amplitude of velocity 
fluctuations. Dotted lines show the function $V(k)$ for the velocity 
dispersion $\sigma_v=0$ and $\sigma_v=600$ km s$^{-1}$. 
The dot-dashed line describes the approximation (22). Filled circles 
demonstrate the velocity rms amplitude from peculiar velocities. 
(b) The rms peculiar velocity of matter for the velocity spectra presented 
in panel (a). The dot-dashed line describes the approximation (24). 
Filled square demonstrates the observed rms cluster peculiar velocity. 
(c) The rms amplitude of the bulk flow. Filled triangles 
demonstrate the observed bulk velocities derived from the peculiar 
velocities.}

\figcaption[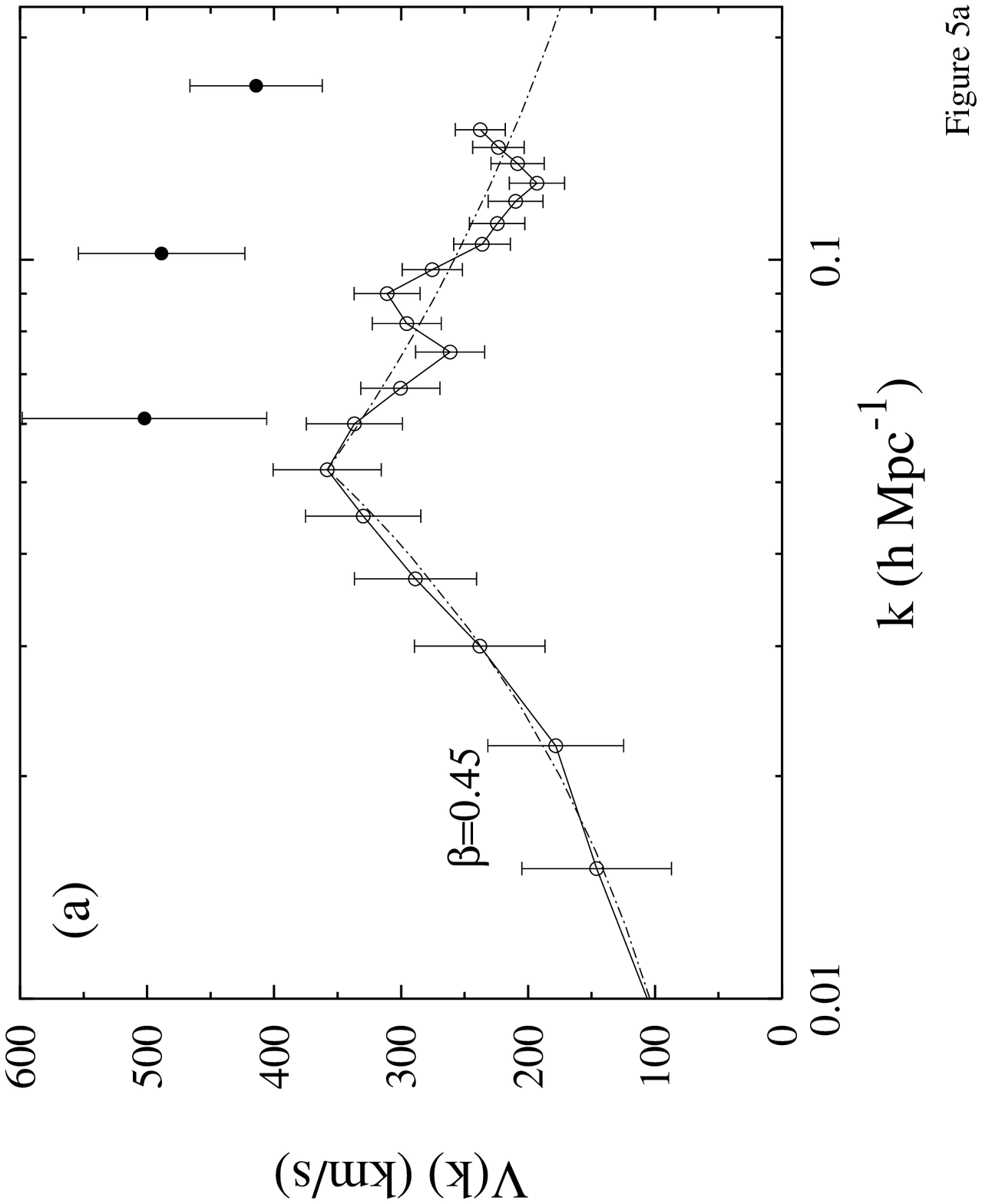,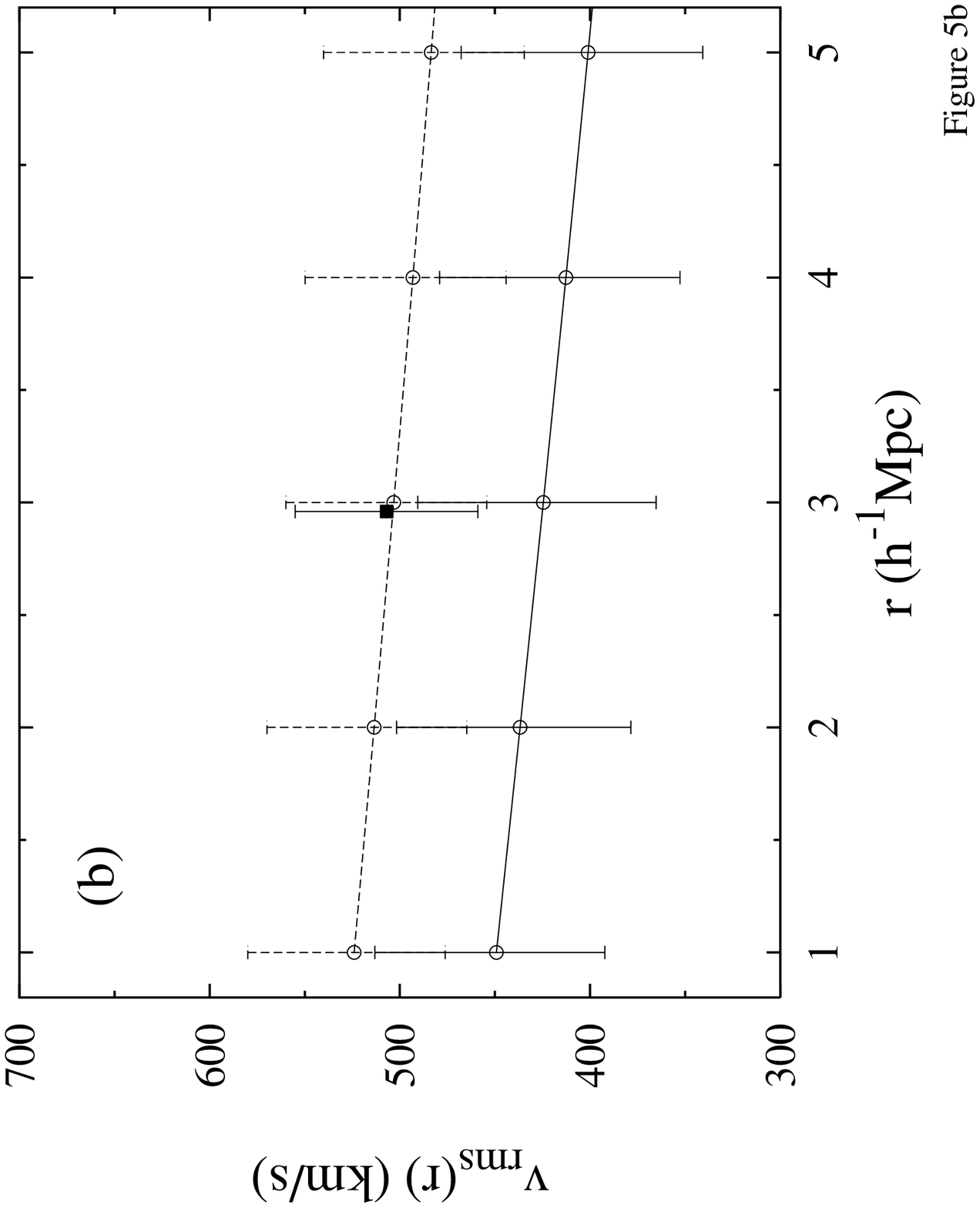,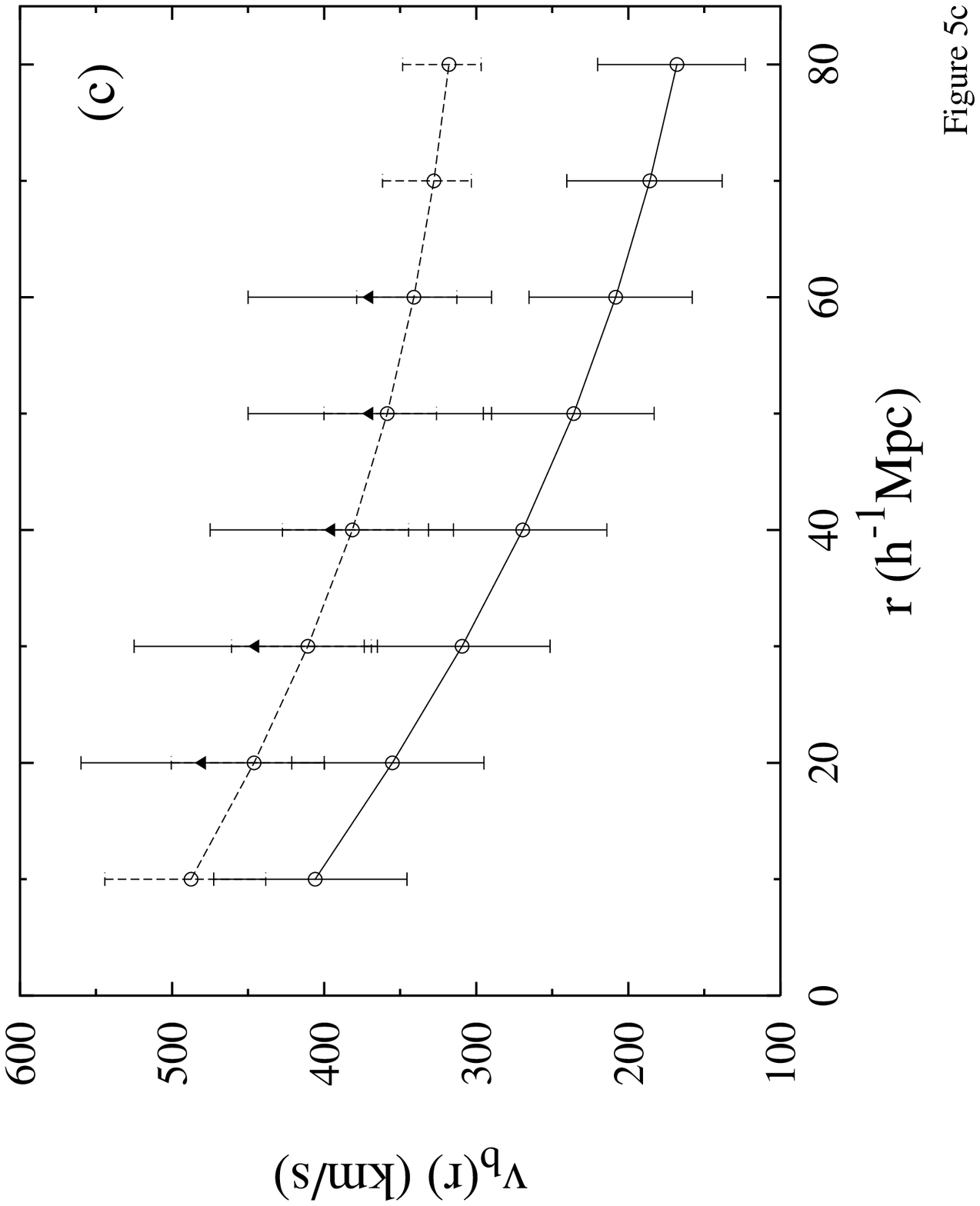] {Velocity fluctuations 
in the Stromlo-APM redshift survey for the parameter $\beta=0.45$. 
(a) The rms amplitude of velocity fluctuations (open circles with 
solid line). 
The dot-dashed line describes the approximation (22). 
Filled circles show the velocity rms amplitude from peculiar velocities. 
(b) The rms peculiar 
velocity of matter for the parameter $v_L=0$ (solid line) and for the 
parameter $v_L=270$ km s$^{-1}$ (dashed line). Filled square demonstrates 
the observed rms cluster peculiar velocity. (c) The rms amplitude of the 
bulk flow for the parameter $v_L=0$ (solid line) and for the parameter
$v_L=270$ km s$^{-1}$ (dashed line). Filled triangles demonstrate the 
observed bulk velocities derived from the peculiar velocities.}

\figcaption[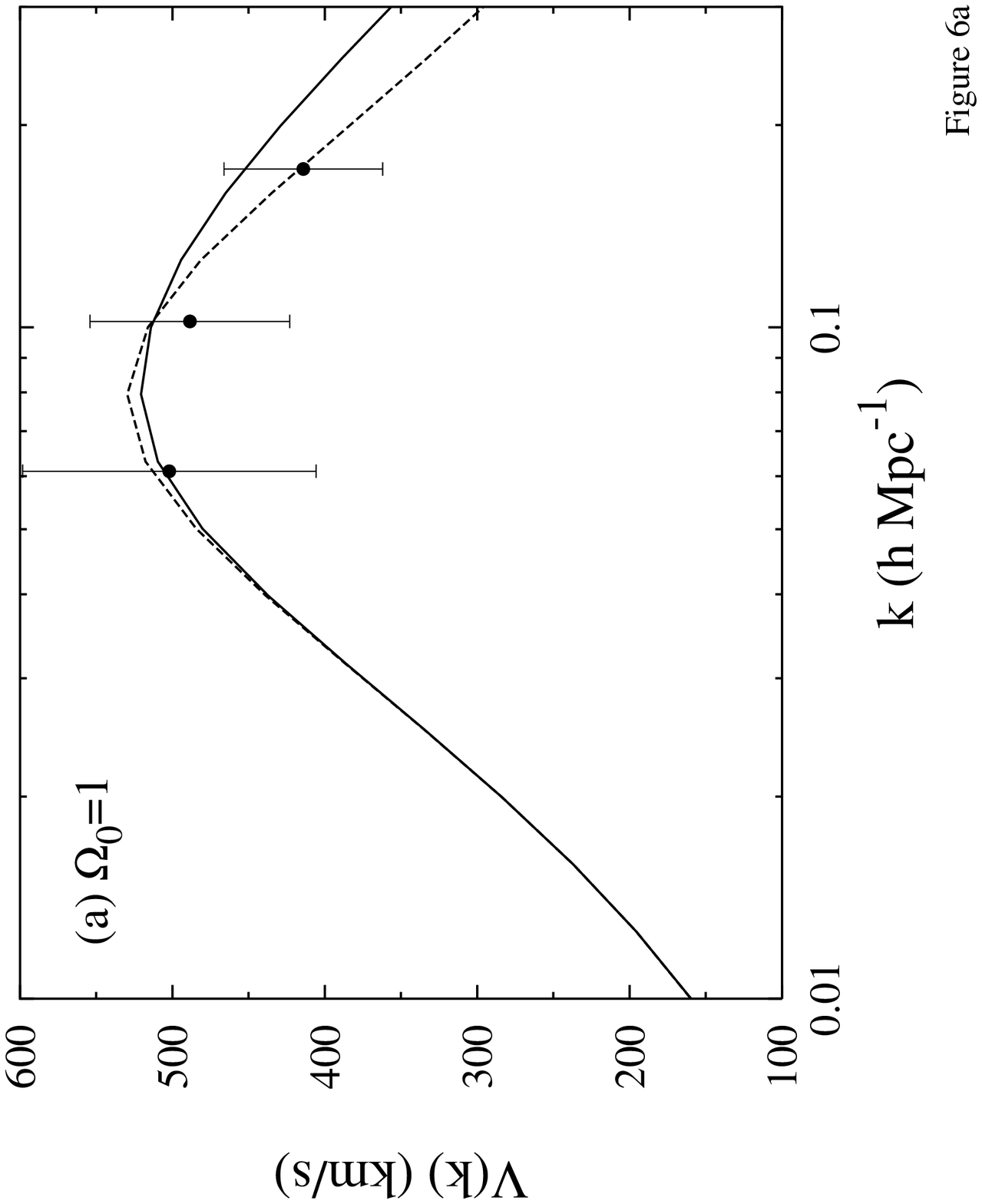,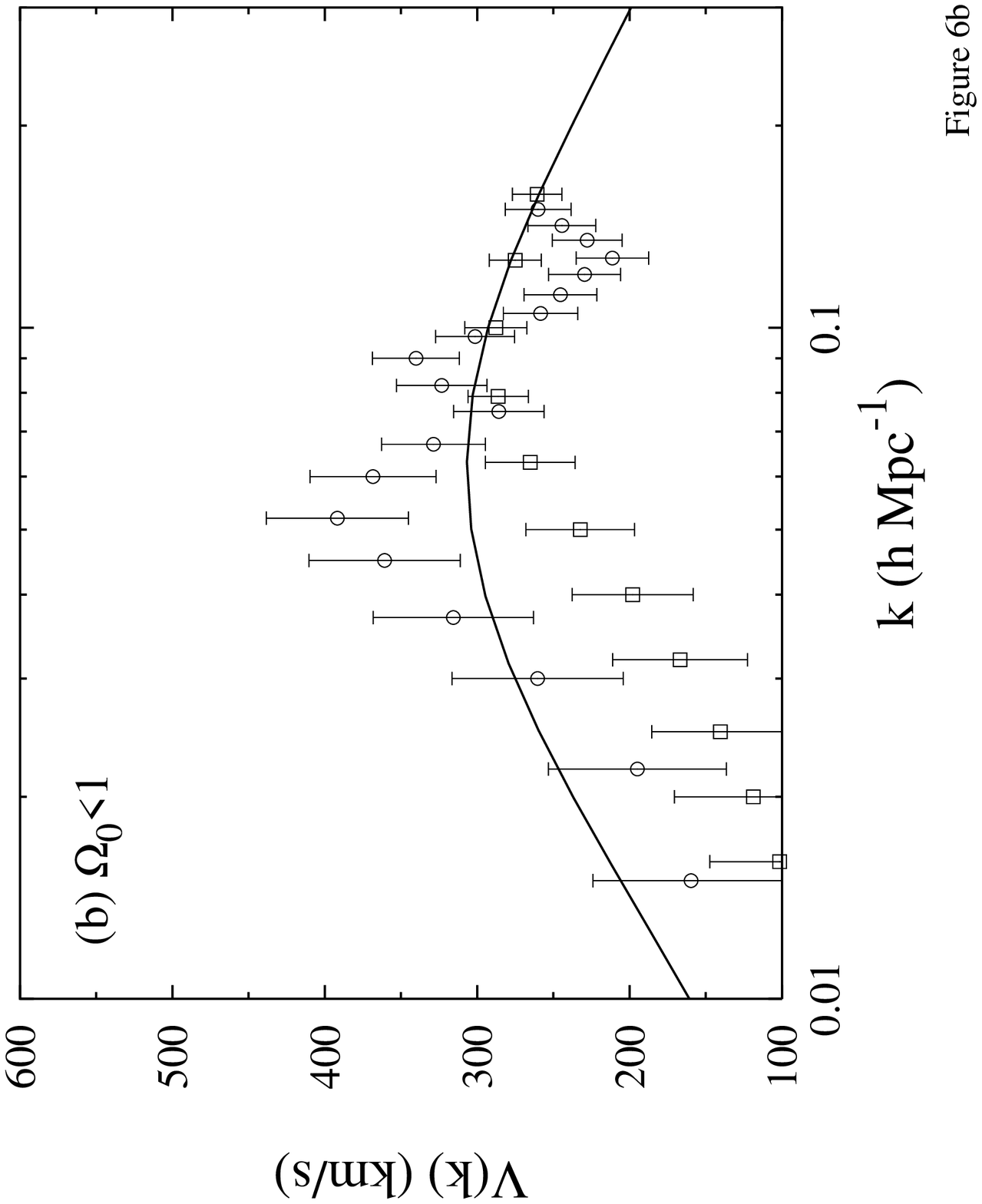] {(a) The rms amplitude of velocity fluctuations 
in the CHDM models with neutrino density $\Omega_{\nu}=0.2$ (solid line) 
and $\Omega_{\nu}=0.3$ (dashed line). Filled circles demonstrate the 
velocity rms amplitude derived from peculiar velocities of galaxies.
(b) The rms amplitude of velocity fluctuations in the flat CDM model
with density parameter $\Omega_0=0.3$. Open circles and squares show
the velocity rms amplitude for the parameter $\beta=0.5$ derived from 
galaxy distribution in the Stromlo-APM and Las Campanas redshift survey,
respectively.}


\clearpage

\begin{references}

Bahcall, N. A., \& Cen, R. 1993, \apj, 407, L49

Bahcall, N. A., Gramann, M., \& Cen, R. 1994, \apj, 436, 23

Bahcall, N. A., \& Oh, S. P. 1996, \apj, 462, L49 

Bertschinger, E., Dekel, A., Faber, S. M., Dressler, A., 
\& Burstein, D. 1990, \apj, 364, 370

Broadhurst, T. J., Ellis, R. S., Koo, D. C., \& Szalay, A. S. 1990,
\nat, 311, 517

Bunn, E. F., \& White, M. 1997, \apj, 480, 6

da Costa, L. N., Vogeley, M. S., Geller, M. J., Huchra, J. P., 
\& Park, C. 1994, \apj, 437, L1

Dekel, A. 1994, \araa, 32, 371

Dekel, A., Burstein, D., \& White, S. D. M. 1996, in {\it Critical 
Dialogues in Cosmology}, ed. Neil Turok (World Scientific, in press) 

Einasto, J., et al. 1997, \nat, 385, 139

Eke, V. R., Cole, S., \& Frenk, C. S. 1996, \mnras, 282, 263

Giovanelli, R., Haynes, M. P., Chamaraux, P., da Costa, L. N.,
Freudling, W., Salzer, J. J., \& Wegner G. 1996, in IAU Symp 168, 
Examining the Big Bang and Diffuse Background Radiation, ed. M. Kafatos
and Y. Kondo (Dordrecht:Kluwer)

Gramann, M., Cen, R., \& Bahcall, N. A. 1993, \apj, 419, 440

Gramann, M, Bahcall, N. A., Cen, R., \& Gott, J. R. 1995, \apj, 441, 449

Gunn, J. E., \& Weinberg, D. H. 1995, in Proc. 35th Herstmonceux Conf.,
Wide-Field Spectroscopy and the Distant Universe, ed. S. J. Maddox \&
A. Aragon-Salamanca (Singapore:World Scientific),3 

Kaiser, N. 1987, \mnras, 227, 1

Kolatt, T. \& Dekel, A. 1997, \apj, 479, 592

Landy, S. D., Shectman, S. A., Lin, H., Kirshner, R. P., Oemler, A., 
\& Tucker, D. 1996, \apj, 456, L1

Lin, H., Kirshner, R. P., Shectman, S. A., Landy, S. D., Oemler, A., 
Tucker, D.L., \& Schechter P. L. 1996, \apj, 471, 617

Loveday, J., Efstathiou, G., Peterson, B. A., \& Maddox, S. J.
1992, \apj, 400, L43

Maddox, S. J., Sutherland, W. J., Efstathiou G., \& Loveday, J. 1990,
\mnras, 243, 692

Marzke, R. O., Geller, M. J., da Costa, L. N., \& Huchra, J. P. 1995,
\aj, 110, 477 

Peacock, J. A., \& Dodds, S. J. 1994, \mnras, 267, 1020

Peebles, P. J. E. 1980, The Large-Scale Structure of the Universe 
(Princeton: Princeton Univ. Press)

Pogosyan, D. Yu., \& Starobinsky, A. A. 1995, \apj, 447, 465

Shectman, S. A., Landy, S. D., Oemler, A., Tucker, D. L., 
Kirshner, R. P., Lin, H., \& Schechter, P. L., 1996, \apj, 470, 172

Starobinsky, A. A., 1992, JETP Lett., 55, 489 

Strauss, M. A., \& Willick, J. A. 1995, \physrep, 261, 271

Tadros, H. \& Efstathiou, G. 1996, \mnras, 282, 1381

White, S. D. M., Efstathiou, G., \& Frenk, C. S., 1993, \mnras, 
262, 1023

Willick, J. A., Courteau, S., Faber, S. M., Burstein, D., Dekel, A., 
\& Strauss, M. A. 1997, \apjs, 109, 333

\end{references}
\end{document}